\documentclass[longauth,]{aa} 

\usepackage{graphicx}
\usepackage[varg]{txfonts}
\usepackage{multirow}
\usepackage{amssymb}
\usepackage{amsmath}

\def\e20{$\times 10^{20}$}
\def\ergsec{erg s$^{-1}$}
\def\kmsec{km s$^{-1}$}
\def\kmsecmpc{km s$^{-1}$ Mpc$^{-1}$}
\def\Mpch{$h^{-1}_{70}~Mpc~$}
\def\ie{i.e.,~}

\def\eg{e.g.,~}

\def\hnot{$H_0$}
\def\kpch{$h^{-1}_{70}~\mathrm{kpc}~$}
\def\kpcharc{$h^{-1}_{70}~\mathrm{kpc}$/\arcsec}
\def\omegaM{$\Omega_\mathrm{M}$}
\def\omegaL{$\Omega_\Lambda$}
\def\today{\number\day -\number\month -\number\year}
\def\Msun{$\mathrm{M}_\odot$}

\newcommand\T{\rule{0pt}{2.6ex}}       
\newcommand\B{\rule[-1.2ex]{0pt}{0pt}} 

\begin{document}

   \title{Intra Cluster Light properties in the CLASH-VLT cluster MACS J1206.2-0847
   \thanks{Based on data collected at the NASJ Subaru telescope, at the ESO VLT
     (prog.ID 186.A-0798), and the NASA HST.}}

   \subtitle{}
%

\author{
 V. Presotto\inst{\ref{MGi},\ref{ABi}}
 \and M. Girardi\inst{\ref{MGi},\ref{ABi}} 
 \and M. Nonino\inst{\ref{ABi}} 
 \and A. Mercurio\inst{\ref{AMe}} 
 \and C. Grillo\inst{\ref{CGr}}
 \and P. Rosati\inst{\ref{PRoNew}}
 \and A. Biviano\inst{\ref{ABi}} 
 \and M. Annunziatella\inst{\ref{MGi},\ref{ABi}}
 \and I. Balestra\inst{\ref{AMe},\ref{ABi}}
 \and W. Cui\inst{\ref{MGi},\ref{ABi},\ref{WCui3}}
 \and B. Sartoris\inst{\ref{MGi},\ref{ABi}\ref{BSar3}}
 \and D. Lemze\inst{\ref{DLe}} 
 \and B. Ascaso\inst{\ref{NBe}} 
 \and J. Moustakas\inst{\ref{JMou}}
 \and H. Ford\inst{\ref{DLe}}
 \and A. Fritz\inst{\ref{MSc}}
 \and O. Czoske\inst{\ref{OCz}}
 \and S. Ettori\inst{\ref{SEt},\ref{MMe}}
 \and U. Kuchner\inst{\ref{OCz}}
 \and M. Lombardi\inst{\ref{MLo}}
 \and C. Maier\inst{\ref{OCz}}
 \and E. Medezinski\inst{\ref{EMe}}
 \and A. Molino\inst{\ref{NBe}}
 \and M. Scodeggio\inst{\ref{MSc}}
 \and V. Strazzullo\inst{\ref{VStr}}
 \and P. Tozzi\inst{\ref{PTo}}
 \and B. Ziegler\inst{\ref{OCz}}
 \and M. Bartelmann\inst{\ref{MBa}}
 \and N. Benitez\inst{\ref{NBe}}
 \and L. Bradley\inst{\ref{MPo}}
 \and M. Brescia\inst{\ref{AMe}}
 \and T. Broadhurst\inst{\ref{TBr}}
 \and D. Coe\inst{\ref{MPo}}
 \and M. Donahue\inst{\ref{MDo}}
 \and R. Gobat\inst{\ref{RGo}} 
 \and G. Graves\inst{\ref{GGr1},\ref{GGr2}}
 \and D. Kelson\inst{\ref{DKe}}
 \and A. Koekemoer\inst{\ref{MPo}}
 \and P. Melchior\inst{\ref{PMe}}
 \and M. Meneghetti\inst{\ref{SEt},\ref{MMe}}
 \and J. Merten\inst{\ref{LMo}}
 \and L. Moustakas\inst{\ref{LMo}}
 \and E. Munari\inst{\ref{MGi},\ref{ABi}}
 \and M. Postman\inst{\ref{MPo}}
 \and E. Reg\H{o}s\inst{\ref{ERe}}
 \and S. Seitz\inst{\ref{SSe1},\ref{SSe2}}
 \and K. Umetsu\inst{\ref{KUm}}
 \and W. Zheng\inst{\ref{DLe}}
 \and A. Zitrin\inst{\ref{AZi},\ref{AZi2}}
}

  \institute{
  Dipartimento di Fisica, Univ. degli Studi di Trieste, via Tiepolo 11, I-34143 Trieste, Italy\label{MGi} \and
  INAF/Osservatorio Astronomico di Trieste, via G. B. Tiepolo 11, I-34131, Trieste, Italy\label{ABi} \and
 INAF/Osservatorio Astronomico di Capodimonte, Via Moiariello 16 I-80131 Napoli, Italy\label{AMe} \and
 Dark Cosmology Centre, Niels Bohr Institute, University of Copenhagen, Juliane Maries Vej 30, 2100 Copenhagen, Denmark\label{CGr} \and
 Dipartimento di Fisica e Scienze della Terra, Universita' di Ferrara, Via Saragat, 1, I-44122, Ferrara, Italy \label{PRoNew} \and
 ICRAR, University of Western Australia, 35 Stirling Highway, Crawley, Western Australia 6009, Australia \label{WCui3} \and
 INFN, Sezione di Trieste, Via Valerio 2, I-34127 Trieste, Italy \label{BSar3} \and
 Department of Physics and Astronomy, The Johns Hopkins University, 3400 North Charles Street, Baltimore, MD 21218, USA\label{DLe} \and
 Instituto de Astrof\'{\i}sica de Andaluc\'{\i}a (CSIC), C/Camino Bajo de Hu\'etor 24, Granada 18008, Spain\label{NBe} \and
 Department of Physics and Astronomy, Siena College, 515 Loudon Road, Loudonville, NY 12211, USA \label{JMou} \and
 INAF/IASF-Milano, via Bassini 15, 20133 Milano, Italy\label{MSc} \and
 University of Vienna, Department of Astrophysics, T\"urkenschanzstr. 17, 1180 Wien, Austria\label{OCz} \and
 INAF/Osservatorio Astronomico di Bologna, via Ranzani 1, I-40127 Bologna, Italy\label{SEt} \and
 INFN, Sezione di Bologna; Via Ranzani 1, I-40127 Bologna, Italy\label{MMe}  \and
 Dipartimento di Fisica, Universitá degli Studi di Milano, via Celoria 16, I-20133 Milan, Italy\label{MLo} \and
 Department of Physics and Astronomy, The Johns Hopkins University, 3400 North Charles Street, Baltimore, MD 21218, USA\label{EMe} \and
 CEA Saclay, Orme des Merisiers, F-91191 Gif sur Yvette, France\label{VStr} \and
 INAF/Osservatorio Astrofisico di Arcetri, Largo E. Fermi 5, 50125 Firenze, Italy\label{PTo} \and
 Institut für Theoretische Astrophysik, Zentrum für Astronomie, Universit\"at Heidelberg, Philosophenweg 12, D-69120 Heidelberg, Germany\label{MBa} \and
 Space Telescope Science Institute, 3700 San Martin Drive, Baltimore, MD 21218, USA\label{MPo} \and
 Department of Theoretical Physics, University of the Basque Country, P. O. Box 644, 48080 Bilbao, Spain\label{TBr} \and
 Department of Physics and Astronomy, Michigan State University, East Lansing, MI 48824, USA\label{MDo} \and
 Laboratoire AIM-Paris-Saclay, CEA/DSM-CNRS, Université Paris Diderot, Irfu/Service d'Astrophysique, CEA Saclay, Orme des Merisiers, F-91191 Gif sur Yvette, France\label{RGo} \and
 Department of Astronomy, University of California, Berkeley, CA, USA\label{GGr1} \and
 Department of Astrophysical Sciences, Princeton University, Princeton, NJ, USA\label{GGr2} \and
 Observatories of the Carnegie Institution of Washington, Pasadena, CA 91 101, USA\label{DKe} \and
 Department of Physics, The Ohio State University, Columbus, OH, USA \label{PMe} \and
 Jet Propulsion Laboratory, California Institute of Technology, 4800 Oak Grove Dr, Pasadena, CA 91109, USA\label{LMo} \and
 European Laboratory for Particle Physics (CERN), CH-1211, Geneva 23, Switzerland\label{ERe} \and
 University Observatory Munich, Scheinerstrasse 1, D-81679 M\"unchen, Germany\label{SSe1} \and
 Max-Planck-Institut f\"ur extraterrestrische Physik, Postfach 1312, Giessenbachstr., D-85741 Garching, Germany \label{SSe2} \and
 Institute of Astronomy and Astrophysics, Academia Sinica, P. O. Box 23-141, Taipei 10617, Taiwan\label{KUm} \and
 Cahill Center for Astronomy and Astrophysics, California Institute of Technology, MS 249-17, Pasadena, CA 91125, USA\label{AZi} \and
 Hubble Fellow\label{AZi2}
}

   \date{}

 
   \abstract
    {} 
    {We aim at constraining the assembly history of clusters by studying the intra cluster light (ICL) properties, 
estimating its contribution to the fraction of baryons in stars, f$_*$, and understanding possible systematics/bias 
using different ICL detection techniques.}
    {We developed an automated method, {\it GALtoICL}, based on the software GALAPAGOS to obtain a refined version of 
typical BCG+ICL maps.
We applied this method to our test case MACS J1206.2-0847, a massive cluster located at z$\sim$0.44, that is part of 
the CLASH sample.
Using deep multi-band SUBARU images, we extracted the surface brightness (SB) profile
of the BCG+ICL and we studied the ICL morphology, color, and contribution to f$_*$ out to R$_{500}$. 
We repeated the same analysis using a different definition of the ICL, {\it SBlimit} method, \ie a SB cut-off level, 
to compare the results.}
    {The most peculiar feature of the ICL in MACS1206 is its asymmetric radial distribution, with an excess in the SE 
direction and extending towards the 2$^{nd}$ brightest cluster galaxy which is a Post Starburst galaxy. This suggests
an interaction between the BCG and this galaxy that dates back to $\tau\leq1.5$Gyr.
The BCG+ICL stellar content is $\sim8$\% of M$_{*,\,500}$ and the (de-) projected baryon fraction in stars is 
 f$_{*}=0.0177 (0.0116)$, in excellent agreement with recent 
results. The {\it SBlimit} method provides systematically higher ICL fractions and this effect is larger at lower SB 
limits. This is due to the light from the outer envelopes of member galaxies that contaminate the ICL. Though more 
time consuming, the {\it GALtoICL} method provides safer ICL detections that are almost free of this contamination.
This is one of the few ICL study at redshift z $>$ 0.3. At completion, the CLASH/VLT program will allow us to 
extend this analysis to a statistically significant cluster sample spanning a wide redshift range: 0.2$\lesssim$z$\lesssim$0.6.}
    {}


\titlerunning{ICL properties in CLASH-VLT cluster MACS1206}
\authorrunning{V. Presotto et al.}

\maketitle
%

\section{Introduction}
\label{Sec:Int}

Since its first discovery by \citet{Zwicky1951} to the most recent works \citep{Guennou2012,Burke2012,Adami2012} the
intra cluster light (ICL) has gained increasing interest because it can help us understanding both 
the assembly history of galaxy clusters and its contribution to the baryonic budget. 
The ICL consists of stars which are bound to the cluster potential after being stripped from 
member galaxies as they interacted and merged with either the brightest cluster galaxy 
(BCG) or the other member galaxies \citep{Murante2004,Sommer2005,Monaco2006,Murante2007,Conroy2007,Puchwein2010,Rudick2011,Cui2013,Contini2013}. 
The ICL signature can be seen in the surface brightness (SB) profile of the BCG as an excess of light with respect to 
the typical r$^{1/4}$ law \citep{DeVaucouleurs1953}. \citet{Gonzalez2005} showed that a double r$^{1/4}$ model 
provides a better fit to the BCG+ICL SB profile and that the ICL has a more concentrated profile than that of the total cluster light
 \citep[see also][]{Zibetti2005}.

The origin of the ICL strictly connects it to the  evolutionary history of the clusters, thus, we can recall the assembly history of 
the clusters by studying the ICL properties. 
The ICL colors can provide us information on the timescales involved in ICL formation and on 
its progenitors when compared to BCG colors.
Some works found that ICL colors are consistent with those of the BCG 
\citep[\eg][]{Zibetti2005,Krick2007,Pierini2008,Rudick2010}, 
suggesting that the ICL has been originated by ongoing interactions among cluster members and the BCG. 
The merging cluster in the sample of \citet{Pierini2008} and some compact groups \citep{DaRocha2005} 
represent an exception showing bluer colors for the ICL, hinting to either {\it in-situ} star formation
or blue dwarf disruption after interaction.

Usually the ICL is found to be strongly aligned with the position angle (PA) of the BCG \citep{Gonzalez2005,Zibetti2005},
but there are cases of misalignment and/or prominent features/plumes \citep{Mihos2005,Krick2007}.
Studying the connections between the ICL spatial distribution and the presence of cluster substructures 
can shed a light on the origin of the ICL and its connection to the assembly history of the cluster. 
ICL plume-like structures bridging together the BCG and other galaxies, arcs and tidal streams of ICL 
have been found by many works \citep[\eg][]{Gregg1998,Calcaneo2000,Feldmeier2004,Krick2006,DaRocha2008}. According to simulations 
these features trace recent interactions and/or merger events between galaxies and/or clusters and they are 
supposed to last only $\sim$1.5 times their dynamical timescale because of disruption by cluster tidal field \citep{Rudick2009}.
\citet{Adami2005,Krick2007} also found an association between ICL sources and infalling groups of galaxies and they
used it to infer the dynamical evolution of the clusters.

Beside characterizing the ICL properties and the specific evolution of a single cluster, the ICL can be put in a much more
comprehensive context by determining its contribution to the total stellar cluster mass and, as a consequence, to
the baryon fraction.
Observational studies show fractions of ICL ranging from few percent of the total light up to half of it 
\citep{Feldmeier2004,DaRocha2005,Zibetti2005,Krick2007,Gonzalez2007,DaRocha2008,Guennou2012,Burke2012,Adami2012}, 
depending on enclosing radius, and/or cluster mass.
On top of this there is no common definition of ICL both among observational works and
simulations. Ideally the ICL consists of the residual light 
after having subtracted the contribution of all galaxies, including the BCG. However both choosing
the separation between the BCG and the ICL, and determining the best fit model of member galaxies is 
a difficult task. As a consequence some studies prefer to focus on a BCG+ICL map and mask other members
\citep{Gonzalez2005,Gonzalez2007}, while other authors chose to mask all galaxies down to different
arbitrary surface brightness levels, \citep{Zibetti2005,Krick2007,Burke2012}, and finally \citet{DaRocha2005,Guennou2012}
remove all the galaxy contribution via a wavelet technique.
Different ICL detection methods can suffer from different systematics/bias thus providing discordant ICL fractions
as shown for simulations \citep{Cui2013}. This variety of ICL definitions can explain part of the lack of a general 
consensus on the effective role played by the ICL in the cluster baryon budget.

Moreover the fraction of ICL can correlate with global cluster properties such as mass,
projected distance and redshift depending on the dominant process and epoch at which they occur 
\citep[see][for a comprehensive description of the origin of these correlations]{Krick2007}. 
\citet{Guennou2012} found only a weak correlation between the ICL content and the cluster velocity dispersion/mass 
and there is no variation in the amount of ICL between z = 0.4 and z = 0.8. The absence or mildness of these trends
is confirmed also at lower redshifts, \ie z $<$ 0.3, \citep{Zibetti2005,Krick2007}. These findings are inconsistent
with most of the previous results from both cosmological and analytical simulations which generally agree 
with an increasing ICL fraction as cluster mass grows \citep{Murante2004,Lin2004,Purcell2007,Watson2012}.
However recent simulations suggest a much weaker dependence of the ICL fraction on cluster mass 
\citep{Murante2007,Dolag2010,Puchwein2010,Martel2012,Cui2013}.

Apparently ICL is a promising and complementary way to understand the 
mechanisms occurring in galaxy cluster and their constituents, however there are two main disadvantages.
First the ICL features typically have extremely faint surface brightnesses of $\sim$1\% of
the brightness of the night sky, making their study extremely
difficult. Secondly, the surface brightness dimming increases with redshift as: $(1 + z)^4$. As a consequence, detecting the ICL is very
difficult and there are only few detections at $z>0.3$ \citep{Jee2010,Guennou2012,Burke2012,Adami2012,Giallongo2013}.

In this paper we present our ICL detection and measurement method and the results we obtained from optical  
images of MACS1206.2-0847 (hereafter MACS1206), one cluster in the Cluster Lensing And Supernova survey with Hubble (CLASH) 
sample \citep{Postman2012}. 
Overall this cluster is one of the most massive, M$_{200}$ = 1.41$\times$10$^{15}$ \Msun, among the CLASH sample and it is located 
at a medium-redshift, z$\sim$0.44, with plenty 
of ancillary information, so it is a suitable case in order to test the performances of our ICL detection method. 
The CLASH survey comprises 25 massive clusters of galaxies 
in the redshift range $0.2 \lesssim z \lesssim 0.9$. Among these, 14 have been selected for spectroscopic follow-up at the VLT.
At completion, both photometric and dynamical properties of each cluster will be available allowing the study of 
ICL and its connection to cluster properties over a wide redshift range. Using deep multi-band images from 
SUBARU, we studied the colors and the morphology of the ICL in MACS1206, as well as its connection to cluster 
substructures and its contribution to the total baryon budget. We then compare these results with those we obtain 
applying different ICL detection methods, in order to explore advantages/disadvantages of each method and to reveal possible 
systematics in each method.

In Sect. \ref{Sec:data} we show the data set we used and the details of the reduction, in Sect. \ref{Sec:method}
we explain our ICL detection and measurement method. Sect. \ref{Sec:results_M1206} describes our results in terms of 
both ICL properties and its contribution to the total cluster light/mass. We discuss our results in Sect. 
\ref{Sec:discussion} and in Sect. \ref{Sec:conclusions} we draw our conclusions and future prospects.

Throughout this paper we use \hnot = 70 \kmsecmpc, \omegaM = 0.3, and \omegaL = 0.7, which gives 5.685 \kpcharc 
at z=0.44, the distance of MACS1206.

\section{Data}
\label{Sec:data}

CLASH is one of the 3 multi-cycle treasury program of HST targeting 25 relaxed galaxy clusters with mass range 
$5-30 \times 10^{14}$ \Msun\, and redshift range $0.2 \lesssim z \lesssim 0.9$ and providing images for each cluster
in 16 pass-bands using WFC3/UVIS, WFC3/IR and ACS/WFC \citep[see][for a detailed description of the survey]{Postman2012}. 
MACS1206 is part of the CLASH sample and it has also been selected for the CLASH-VLT follow-up proposal \citep{Rosati2013} 
and for SUBARU imaging for the weak lensing program \citep{Umetsu2012}. 
We choose this cluster as the test case for our analysis because it is the first cluster for which VLT data reduction
is completed, thus we have a wealth of both photometric and 
spectroscopic information. In this Section we describe the data set at our disposal and the reduction techniques. 

\begin{table}
\caption{Photometric data set summary.}
\label{tab:dataset}
\centering

\begin{tabular}{l c c c}
\hline\hline

\noalign{\smallskip}
\multicolumn{4}{c}{SUBARU data} \\
\noalign{\smallskip}
\hline

 Filter &  exposure time & seeing & Mag lim  \\
 & (ks) & (\arcsec) &  (AB mag) \\
\hline

B & 2.4 & 1.01 & 26.5 \\
V & 2.2 & 0.95 & 26.5 \\
Rc & 2.9 & 0.78 & 26.2 \\
Ic & 3.6 & 0.71 & 26.0 \\
z' & 1.6 & 0.58 & 25.0 \\

\hline\hline

\end{tabular} 

\end{table}

\subsection{Photometry}
\label{Subsec:photometry}

We analyzed deep BVRcIcz images obtained with the Suprime-Cam mounted at SUBARU telescope and that are available in the Subaru archive, 
SMOKA\footnote{http://smoka.nao.ac.jp}. A full description of 
the observations can be found in \citet{Umetsu2012} while for a detailed explanation of data reduction
we refer the reader to \citet{Nonino2009}, here we only provide a brief description.
The typical seeing in the final sky subtracted images varies from 0.58\arcsec in the z band up to
1.01\arcsec in the B band with exposure times ranging between 1.6 ks and 3.6 ks with a pixel scale of 0.2 
\arcsec pixel$^{-1}$. The limiting magnitudes are m$_B$ = 26.5, m$_V$ = 26.5, m$_{Rc}$ = 26.2, m$_{Ic}$ = 26.0,
and m$_z$ = 25.0 mag for a 3$\sigma$ limiting detection within a 2\arcsec diameter aperture, see Tab. \ref{tab:dataset} 
for a summary of our photometric data set. 

Sky subtraction and diffuse low-level light patterns removal are crucial because part of the ICL can be removed in these steps
of the data reduction. As described in \citet{Nonino2009}, we carefully determine the background by a back-and-forth process. First, we 
detect sources in a preliminary stacked image, the area covered by each source is enlarged by 20\%, and the corresponding segmentation 
map is used to flag the same pixels in each original image. Flagged pixels in each individual image are replaced by a random value 
normally distributed with mean and standard deviation obtained by a $\sim30\arcsec\times30\arcsec$ box surrounding each pixel, excluding 
flagged pixel values. Finally,
each resulting image is wavelet transformed and the background of each image corresponds to the lowest order plane of the wavelet 
transformation. To ensure that this process does not affect our estimation of the ICL we use our BCG+ICL map of MACS1206, see 
Sects. \ref{Subsec:method_desc} and \ref{Sec:results_M1206}, as a control map. Only 0.37\% of the BCG+ICL map pixels having a value larger than 3$\times\sigma_{sky}$ 
fall out of the enlarged segmentation map, where $\sigma_{sky}$ refers to the $\sigma$ of the residuals after sky subtraction as 
estimated in an area free from any source contamination.
None of these pixels is recognized as a source by SExtractor, \ie these few pixels are 
randomly distributed and they most probably represent fluctuations. If we restrict this analysis to a 3\arcmin$\times$2\arcmin area 
surrounding the BCG, then the percentage of outlier pixels decreases to 0.09\%. Thus, the enlarged mask used in the background 
subtraction process ensures us that no pixels associated to the ICL has been oversubtracted. As a consequence, background subtraction 
does not affect our ICL estimation and we consider $\sigma_{sky}$ as our limit to detect the ICL. As a further check, we applied the 
{\it SBlimit} method, see Sect. \ref{Subsec:ICL_cfr_SBlim}, to the F625W HST stacked image, \ie the closest HST filter to the Rc SUBARU band, 
and we cross-correlate it with the corresponding Rc band image. This way we can check whether the spatial distribution of the ICL down to different SB levels is the same 
in both images. According to the cross-correlation analysis, the optimal x,y shift to match the two images is zero for all the SB 
levels. Given that the HST image has been reduced in an indipendent way, \ie using a different background subtraction process, this 
ensures us that we did not remove any real low surface brightness sources during the data reduction.

The stellar point spread functions (PSFs) were measured from a combination of unsaturated stars 
with S/N $\geq 50$ and ellipticity $\leq 0.1$, here ellipticity is defined as (1 - a)/(1 + a), where 
a is the source aspect ratio, \ie an ellipticity of 0.1 corresponds roughly to an aspect ratio of $\sim 0.8$.
The point sources are detected and modeled using SExtractor and PsfeX softwares \citep{Bertin1996,Bertin2011} and their
PSF model is derived solely from the robust combination of their resampled input vignettes.
In the following analysis this PSF model is convolved with the best fit model of each galaxy obtained as described
in Sect. \ref{Subsec:method_desc}.

The B and Rc broad-band filters
nicely probe the spectral region across the 4000 \AA\, break at the cluster redshift, thus the (B-Rc)
color is a good indicator of the galaxy average star formation (SF) history and it can constrain
the characteristics of the bulk of its stellar population. We will use this color to infer information on
the ICL properties.

We obtained magnitudes in each band and the relative colors for all detected sources, these data
were used to derive photometric redshifts, z$_{photo}$, using a method based on neural networks:
The Multi Layer Perceptron with Quasi Newton Algorithm (MLPQNA) \citep{Brescia2013}. 
This method was calibrated on a subsample of objects with spectroscopic redshifts and it
was applied to the whole data-set with available and reliable BVRcIcz band magnitudes
down to m$_{Rc}$ = 25.0 \citep[see][for a detailed description on the z$_{photo}$ estimation]{Biviano2013,Mercurio2013}.
The validation process with spectroscopically measured redshifts makes the estimated z$_{photo}$ 
insensitive to photometric systematic errors
and more robust than methods based on Spectral Energy Distribution (SED) fitting because 
the neural network method do not depend neither on synthesis models nor on photometric zero point accuracy.
Tests on the MLPQNA based on a combination of parameters from different surveys estimate an excellent
accuracy of $\Delta_{\mathrm{z_{photo}}}=0.004\times(1+z_{\rm spec})$ \citep{Cavuoti2012,Brescia2013}.

\subsection{Spectroscopy}
\label{Subsec:spectroscopy}

Though our work is based on the imaging data described in the previous Section, we will also take advantage of the information
from the spectroscopic dataset of CLASH/VLT to interpret our results. Here we only give the basic description of this dataset and 
we refer the reader to \citet{Rosati2013} and references therein for the details.
The CLASH/VLT program is VLT/VIMOS follow-up of 12/25 CLASH clusters, it comprises a total of 98 pointings that were obtained in the 
spectral range of 3700-97000 \AA\, using the medium resolution (MR) and low resolution (LR) grisms, yielding spectral resolutions of 
580 and 180, respectively. 

In the case of MACS1206 12 masks (4 MR, 8 LR) were observed for a total exposure time of 10.7 hours.
Additional spectra were obtained at VLT/FORS2, Magellan telescope, and from literature/archival data \citep{Lamareille2006,Jones2004,Ebeling2009}. 
The final data-set contains 2749 objects with reliable redshift
estimates, z$_{spec}$ with an average error of 75 and 153 \kmsec for spectra in MR and LR mode respectively.

We measure the main spectral features in the observed spectral range, \ie Dn(4000), H$\delta$, [OII], OIII, and H$\alpha$.
Joining this information to the (B-Rc) color allows us to classify each source according to its stellar population 
\citep[see][]{Mercurio2004}. In particular, two classes of galaxies will be relevant for discussing the results we obtained (see 
Sect. \ref{Sec:discussion}):

\begin{enumerate}
 \item {\it Passive galaxies}: sources with Dn(4000) $>$ 1.45 and EW(H$\delta$) < 3.0 \AA;
 \item {\it Red H$\delta$}: sources with Dn(4000) > 1.45 and EW(H$\delta$) > 3.0 \AA.
\end{enumerate}

\begin{figure}
\centering
   \resizebox{\hsize}{!}{\includegraphics[bb=151 0 1342 662, clip]{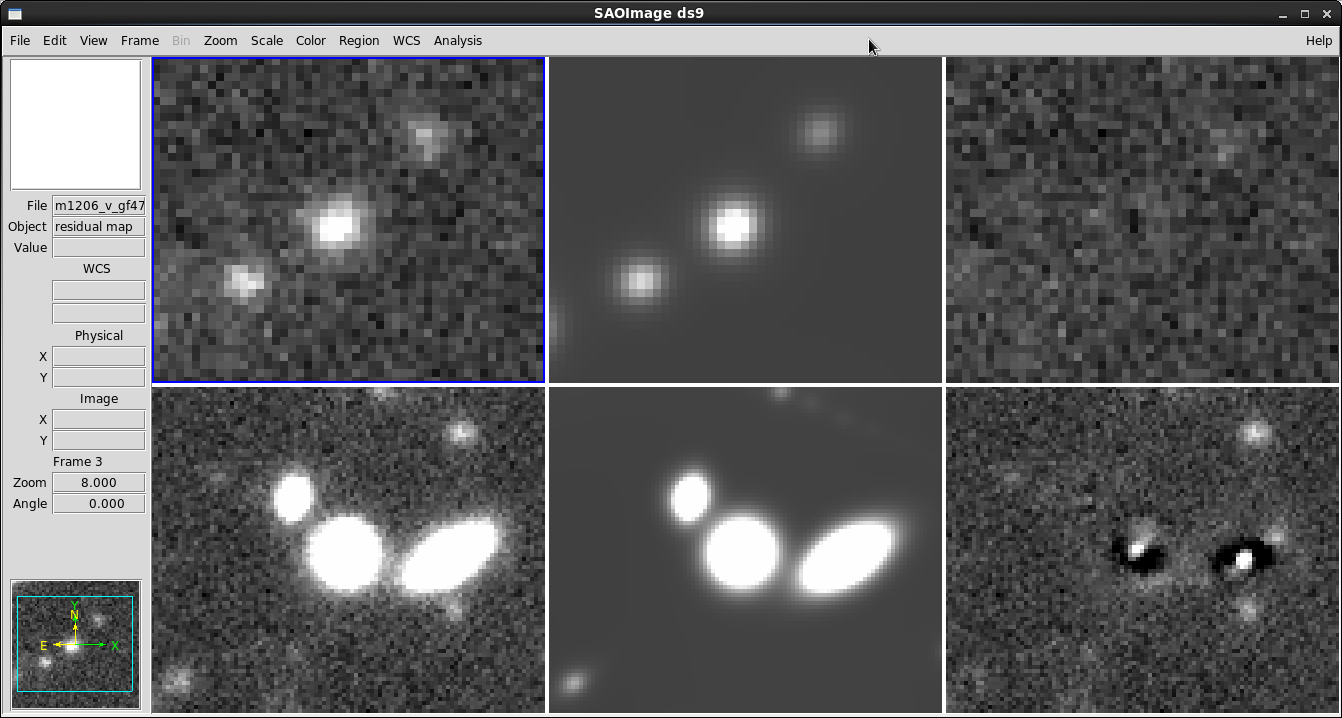}}
   \caption{GALFIT residuals examples. From left to right: original image, best fit model and residuals. 
Top panels refer to a clean fit case, while bottom panels show a case with a high percentage of high residuals.}
      \label{Fig:GALFIT_example}
\end{figure}

\subsection{Cluster membership}
\label{Subsec:cl_membership}

We will need to discriminate between cluster members and fore- back-ground sources both in the ICL 
detection method for MACS1206 (see Sect. \ref{Sec:results_M1206}), and when determining the cluster total light (see 
Sect. \ref{Subsec:ICLfrac}). Photometric information is complementary to the spectroscopic one, thus 
allowing a cluster member association complete down to m$_{Rc}$=25.

The cluster membership for each object is assigned according to 
its spectroscopic redshift, when available, or to its photometric redshift combined
with a color-color cut. We refer the reader to \citet{Biviano2013} for a detailed description
of membership assignment, here we summarize the main steps. In brief, spectroscopic 
members with $18\leq\,m_{R}\,\leq23$ were defined according to the Peak+Gap (P+G) method 
of \citet{Fadda1996}. Photometric members were selected among all the sources 
having a photometric redshift in the range 0.34 $\leq$ z$_{spec/photo}$ $\leq$ 0.54 
and satisfying one of the following color-color cut in the (B-V) and (Rc-Ic) 
diagram:

\begin{eqnarray} 
\mbox{if \,} 0.20 < (B-V) < 0.45 \mbox{\, then:} \nonumber \\
   -0.09 + 0.52 \cdot (B-V) < (Rc-Ic) < 0.21 + 0.52 \cdot (B-V)\\
\mbox{if \,} 0.45 < (B-V) < 0.80 \mbox{\, then:} \nonumber \\
   -0.09 + 0.52 \cdot (B-V) < (Rc-Ic) < 0.36 + 0.52 \cdot (B-V)\\
\mbox{if \,} 0.80 < (B-V) < 1.30 \mbox{\, then:} \nonumber  \\
    0.01 + 0.52 \cdot (B-V) < (Rc-Ic) < 0.36 + 0.52 \cdot (B-V) 
\end{eqnarray}

\section{ICL detection}
\label{Sec:method}

As already mentioned, the ICL consists of the residual light after having removed all the light contribution of galaxies.
Ideally, this can be obtained by subtracting each galaxy best fitting model, choosing among many 
different light profiles, \eg de Vaucouleurs, S{\'e}rsic \citep{Sersic1963,Sersic1968}, Exponential disk, and any combination of them. 
Unfortunately it is not always possible to perfectly fit the 
galaxies, such that the final residuals are not artifacts due to a bad subtraction. As a consequence, most works 
favor masking galaxies down to an arbitrary surface brightness level or subtract a direct image via wavelet 
transformation. In our approach we both subtract the best fit model and mask whenever the fit is not satisfying.

\begin{figure}
\centering
\includegraphics[width=\hsize]{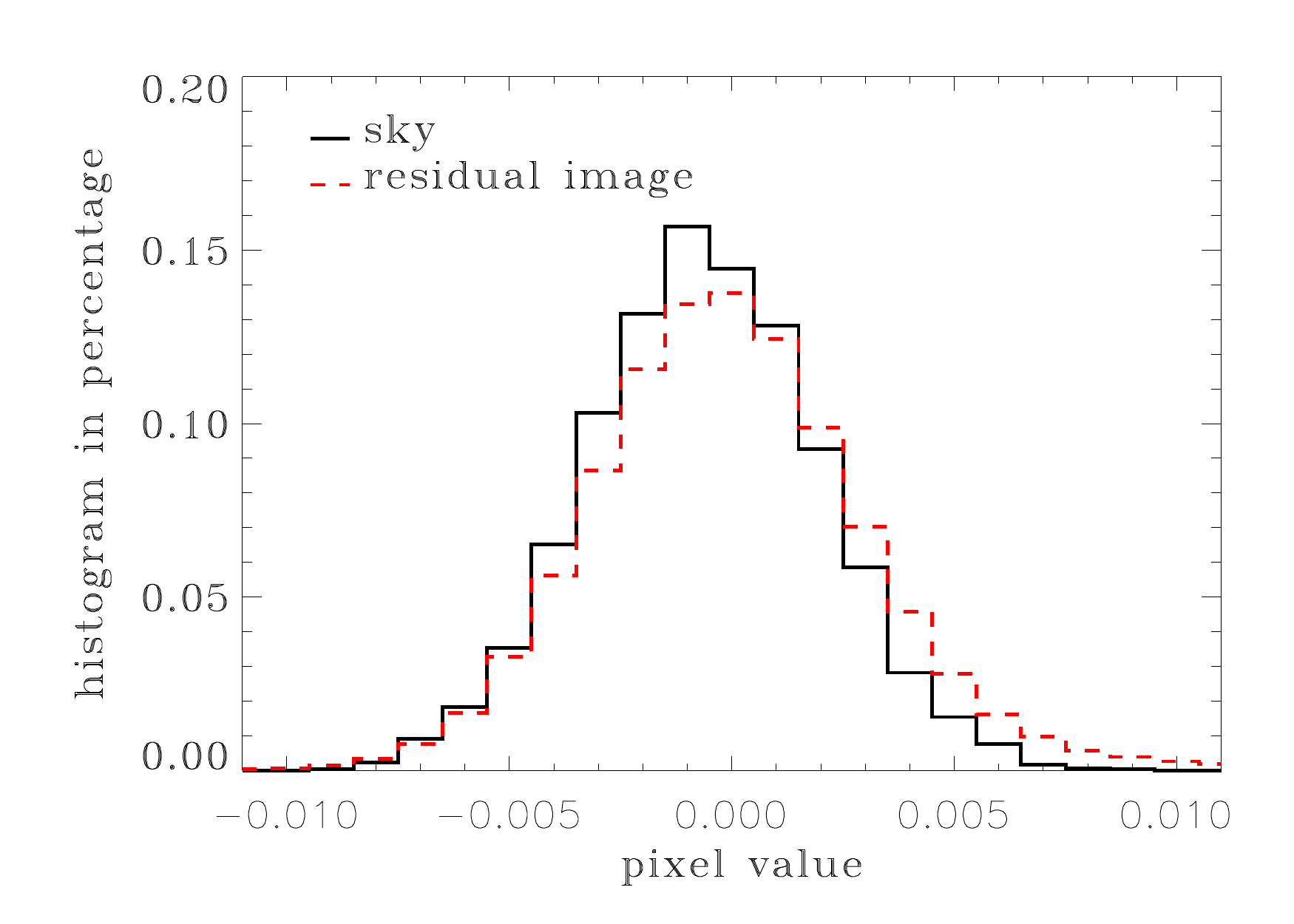}
   \caption{Comparison of pixel values distribution in the residual image (red dashed line) with that of an empty area 
(black solid line), \ie free from source contamination, to identify deviant pixel/sources, see text for details.}
     \label{Fig:cfr_sky_residuals}
\end{figure}

\subsection{Method}
\label{Subsec:method_desc}

We developed an automated method based on the software GALAPAGOS \citep{Barden2012} which makes extensive use of the 
code GALFIT \citep{Peng2010}. GALAPAGOS detects sources in the target image using SExtractor, 
estimates sky background, creates postage stamp images for all detected sources, prepares object masks and finally
performs S{\'e}rsic fitting with GALFIT. We refer the reader to \citet{Barden2012} for more details, here we focus only on 
those steps which are of key importance for our goal. The source detection is performed with a double pass of SExtractor,
one for the bright sources and the second for the faintest ones, then the code recognize whether to discard or to keep a faint 
source depending on its position with respect to the nearest bright source. This minimizes the number of missing/mistaken
faint sources. 

We set the startup parameter file in order to extract faint source with at least S/N 
$\geq$ 1$\sigma_{sky}$. 
We removed the sky background estimation step as we worked with sky subtracted images, however, if this step is 
included, the sky is generally estimated as 0.000 $\pm$ 0.001, this support the goodness of our global sky 
subtraction. The most important step of this code is the postage stamps creation: in this step GALAPAGOS centers the
image section on the source of prime interest and optimizes the area in order to include also the neighbour galaxies. 
This enables GALFIT to simultaneously fit all sources that contribute to the total light in each section, thus 
providing a better fit of each contributing source and removing light coming from the outer envelopes of close 
companions. This cleans the final residual image and ideally provide us the light contribution coming only from ICL. 
It is worth noticing that GALAPAGOS forces GALFIT to fit a single S{\'e}rsic model to each source. The initial guess
for the S{\'e}rsic model parameters correspond to the SExtractor estimates of {\it x\_image, y\_image, mag\_best, {\rm f}(flux\_radius), 
{\rm and} theta\_image}. In many cases a single S{\'e}rsic model is a good approximation but sometimes it can represent a poor 
fit, as described in the following. 
As a last step the code creates the final output catalog containing both SExtractor and GALFIT information for each source. 

At this point we developed an IDL code, {\it GALtoICL}, able to go the other way around: from single postage stamps to a 
final global residual image which we call the BCG+ICL map. The code is composed of 4 main steps:

\begin{enumerate}
  \item creation of a global GALFIT parameter files for a 1000x1000 pixels section of the global science image;
  \item creation of the global best fit model image and the residual image; 
  \item extraction of those sources with a high percentage of high residuals and manual intervention;
  \item creation of the final BCG+ICL map.
\end{enumerate}

At first all sources are listed according to their $\chi ^2$ and their best fit model parameters are stored. 
Then a number of GALFIT set-up files containing at most 50 sources each are created till accounting for all sources 
filling the 1000x1000 section, i.e. $\sim$1150x1150 \kpch at MACS1206 redshift. 
The choice of 50 sources to be modeled in a 1000x1000 pixels section corresponds to the
best compromise of N$_{gals}$ and area that GALFIT is able to deal with due to memory issues. 
All parameters of each source profile are kept fixed as they correspond to their best fit model and we run GALFIT in
{\it model mode}, \ie no fitting, only model image creation based on input parameters. To check whether our conversion 
from (x,y) postage coordinates to (X,Y) global coordinates is well determined we made some tests allowing
(X,Y) to vary within $\pm$ 2 pixels to account for possible errors in centering the sources. We do not find the need for
 any (X,Y) marginal correction and thus we rely on our coordinates transformation. 

Then, all models in each 1000x1000 pixels section are put together to obtain the final global best fit model which is 
then subtracted to the original global science image to obtain the global residual image. 
Bright stars are excluded from the global fit because they might show strong residuals in case of saturation and they need 
specific masking. The code allows you to interactively check the global best fit model, and the residuals images using 
DS9, to update the global best fit model if necessary, and to run again GALFIT. This is the only 
step at which manual intervention is possible. The reason for it is well explained in Fig. \ref{Fig:GALFIT_example} 
where we show two examples of GALFIT performances on postage stamps: from left to right we show the original image, 
its best fit model and the residuals. Top panels refer to a a clean fit case, while bottom panels show a case with a 
high percentage of high residuals. Most of the times we get large residuals because a single S{\'e}rsic model is not enough to 
properly describe the galaxies and more components are needed. 

To identify in an automated way the sources with bad fitting residuals, we compare the distribution of pixels values 
in a region of pure sky with that of the residual image. Fig. \ref{Fig:cfr_sky_residuals} shows these distributions 
with a black solid and red dashed line respectively. Those pixels deviating more than 1, 2, 3, 4, and 5$\sigma_{sky}$ are flagged and through 
SExtractor segmentations maps are connected to the source they belong to. At this point one can choose either to simply
mask them or to perform manual fitting, to up-date the model and to re-run GALFIT to create a better global best fit model 
and residuals images. As a final step the code allows to add ad-hoc masks to those automatically created to fix bad 
pixels, \ie bright saturated stars, spikes. The code is meant to provide BCG+ICL maps, \ie it doesn't create the best fit model 
of the BCG, though one can also choose to obtain only ''ICL'' maps, \ie subtracting also the BCG best fit model.

Once the final model is achieved the code outputs:

\begin{enumerate}
 \item final global best fit model image;
 \item final residual image;
 \item IDs list of deviant sources;
 \item mask images;
 \item final BCG+ICL map with the deviating pixels masked at 1, 2, 3, 4, and 5$\sigma_{sky}$ levels.
\end{enumerate}

The whole process, GALAPAGOS+{\it GALtoICL}, can be iterated twice in order to identify the bright and well deblended 
sources at first and secondly to model also those very faint sources, especially the faint/small satellites of the BCG.
To do this, one can choose the ''ICL'' maps mode and feed again GALAPAGOS with them.

The parameters of the global best fit model can be used as a benchmark for other observed bands by 
running each 50-sources GALFIT set-up file in {\it optimize mode}, \ie allowing X, Y, R$_e$ and Mag 
to change within a certain range.

\begin{figure}
\centering
\includegraphics[width=\hsize]{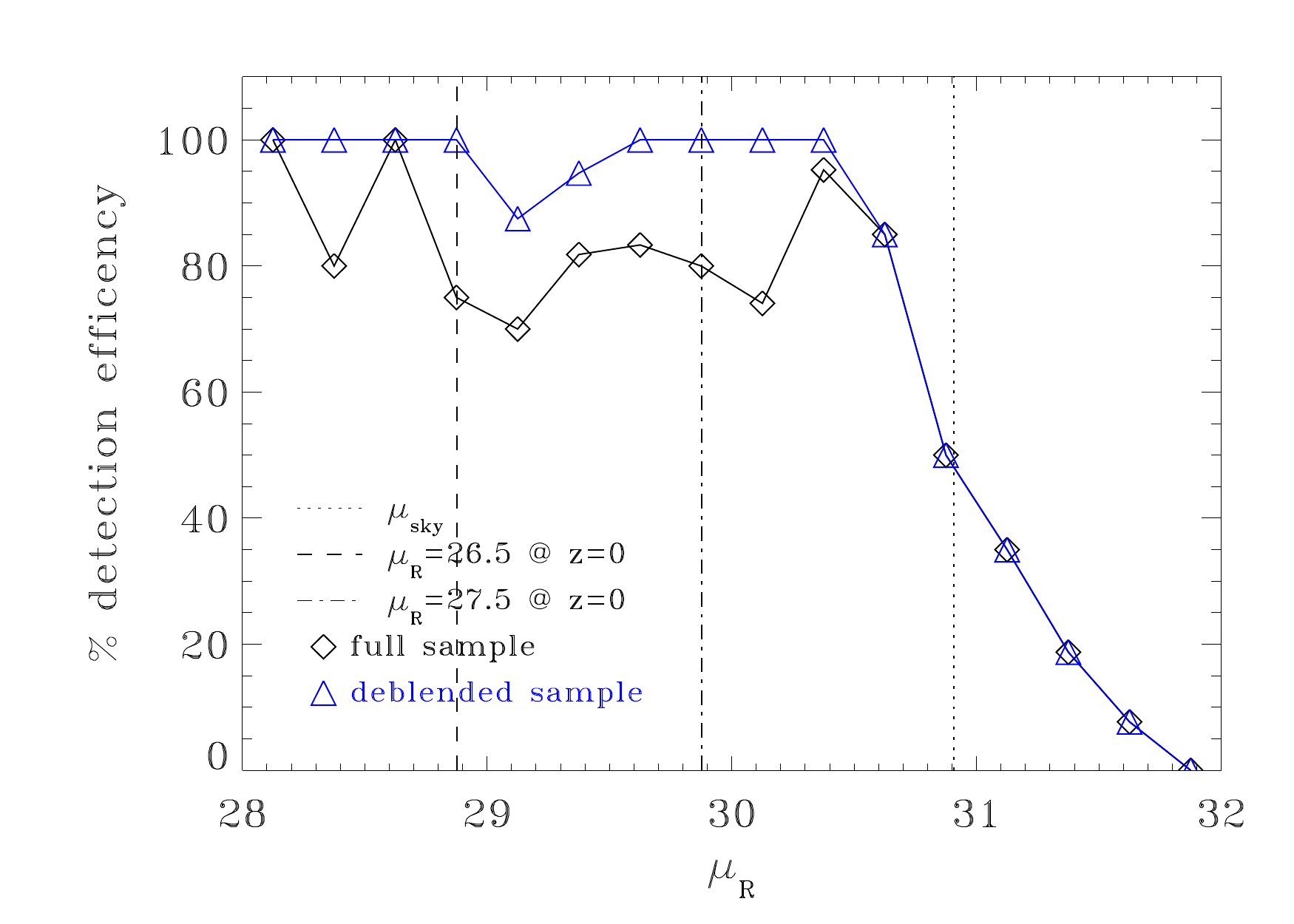}
   \caption{SExtractor detection efficiency as a function of Rc-band surface 
brightness magnitude. Black diamonds refer to the complete sample of fake 
faint sources while blue triangles refer to deblended sources, see text for 
details. The dotted line corresponds to the sky surface brightness while 
dashed and dot-dashed lines correspond to the surface brightness limits 
$\mu_V (z=0) = 26.5;\,27.5$ after accounting for both surface brightness 
dimming and k-correction to transform them in Rc-band limits, see text for details.}
      \label{Fig:detect_efficency}
\end{figure}

\subsection{Detection Efficiency}
\label{Subsec:efficency}

Before applying our detection method to the real images, we test its efficiency
in detecting faint diffuse-light sources. We generate fake faint sources with 
different surface brightnesses and we randomly introduce them into our real Rc-band
images. We also want to determine our ability to deblend and identify these 
faint sources from close bright companions, thus a small percentage of these 
fake sources are forced to lie close to a bright one. We then run our code on 
these real+simulated images.

The artificial faint sources are modeled as de Vaucouleurs profiles with
total magnitude ranging from 21.5 to 24.5 and effective radius varying 
from 20 to 60 pixels, \ie $\sim$ 25-70 \kpch at z=0.44, the cluster MACS1206 
redshift. These parameters choice translates into surface brightness values 
ranging between 28 and 32 mag/arcsec$^{2}$ within a 2\arcsec diameter 
aperture (28 and 30 mag/arcsec$^{2}$ for the blended sources). 
In the Local Universe the ICL is usually detected in the V-band, 
as the light surviving a surface brightness level cut-off, typically 
$\mu_V = 26.5, 27.5$ mag/arcsec$^{2}$ \citep{Feldmeier2004,Mihos2005,Krick2007}.
To compare our results with these studies
we transform these V-band SB levels to the corresponding ones at z=0.44 in 
the Rc-band, \ie we add the surface brightness cosmological dimming 
$2.5\cdot\log(1 + z)^4$ and we applied the k-correction for different bands. 
The latter term is determined running the GALAXEV code on stellar population synthesis 
models \citep{Bruzual2003} for a solar metallicity 
with formation redshift zf=3, a Chabrier initial mass function (IMF)
\citep{Chabrier2003}, and accounting for the stellar population evolution.
Metallicity and formation redshift values are chosen according to
the similarity between typical ICL colors and those of the BCGs
\citep{Zibetti2005,Krick2007,Pierini2008,Rudick2010}.
The resulting SB levels are $\mu_{Rc}(z=0.44) = 28.87, 29.87$ 
mag/arcsec$^{2}$ respectively while our $1\sigma_{sky}$ level corresponds to $\mu_{1\sigma_{sky}} = 30.9$ mag/arcsec$^{2}$,
thus our Rc-band images are deep enough to detect typical diffuse light sources 
redshifted to the considered cluster distance.

\begin{figure*}
\centering
\includegraphics[width=\hsize, bb=1 300 914 552]{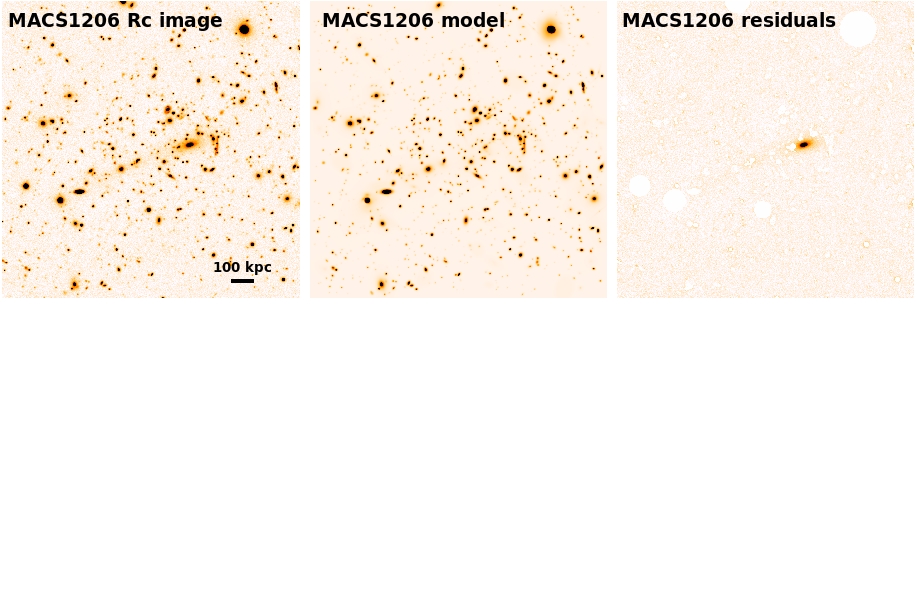}
   \caption{The Rc band image of the MACS1206 core (left panel), its global best fit model 
(central panel) and the final BCG+ICL map masked down to 1$\sigma_{sky}$ level}
      \label{Fig:GALtoICL_results}
\end{figure*}

In Fig. \ref{Fig:detect_efficency} we show our results in terms of SExtractor
detection efficiency as a function of the Rc-band surface brightness.
We set up the SExtractor parameter such that a minimum significant area of 5 pixels
for a 1.5$\sigma$ detection threshold is requested.
Black diamonds refer to the complete sample of artificial
faint sources, \ie both the randomly positioned ones and those lying
close to bright companions, while blue triangles refer only to the well 
deblended sources. 
The dotted line corresponds to the $1\sigma_{sky}$ surface brightness while 
dashed and dot-dashed lines correspond to the surface brightness limits 
$\mu_{V}(z=0) = 26.5;\,27.5$ transformed into the corresponding Rc-band 
value at z=0.44.

We note that the detection efficiency for the deblended sample is 100\% at SB values
well far beyond the lowest $\mu_V(z=0)$ SB level, moreover the detection efficiency at
sky level is almost 50\%. If we consider only the range of SB for which we have 
also blended sources, then the detection efficiency is still more than 70\%.

These tests ensure us that the combination of these deep SUBARU images and
our detection method is good enough to allow diffuse light source 
detections for our test case cluster MACS1206. 

The efficency in recovering the initial parameters, such as R$_e$, S{\'e}rsic index, PA, 
and ellipticity, should be also tested. We used our sample of artificial sources to estimate our 
ability to recover the original parameter value as a function of the surface brightness as measured
 within a 2\arcsec diameter aperture. We split our sample in two subsets: $\mu_{Rc,2\arcsec ap} \leq 26.5$
and $26.5 < \mu_{Rc,2\arcsec ap} < 30.5$ in order to highlight the presence of trends with the SB, if any.
Table \ref{tab:param_retrival} summarizes our results in terms of the median, low and high quartile of the distribution of
either the difference or the ratio between the retrieved and the original parameters for each sub sample.
We do not find any strong trend of the median value as a function of SB, while the errors on the median 
value tend to increase as we move from high to low surface brightness sources. 
This result is in good agreement with \citet{Barden2012} where they used a larger sample of simulated data set-up, 
\ie $\sim10^3$ more galaxies, in order to achieve 
enough statistical significance and to test the recoverability with GALAPAGOS of source parameters and 
its dependence on neighbouring. \citet{Barden2012} showed that GALAPAGOS has optimal performances for
bright galaxies, \ie $\mu_{input}\leq22.5$, while its efficiency decreases at faint magnitudes, \ie $\mu_{input}>22.5$, 
and high S\'ersic indices, \ie $2.5<n<8.0$, see the left panel of their Fig. 14. 
Generally speaking there is no systematic trend/bias for the 
mean recovered parameter value, while the accuracy gets worse from bright to faint sources. 
As far as the influence of neighbouring galaxies is concerned, \citet{Barden2012} showed that GALAPAGOS results do not depend on either the 
magnitude of or the distance from the next neighbour, see their Fig. 16. Given the agreement on parameters retrival tests, we 
did not repeat this test and we rely on their conclusions.

Both the absence of systematic trends and the satisying accuracy level ensure us that the recovered global
model will not be significantly affected by our parameters retrival ability.

\begin{table*}
 \caption{Initial parameter retrival capability of GALAPAGOS+GALtoICL. We report the median value of the distribution of 
either the difference or ratio between the retrieved parameter, Galapagos+GALtoICL (G+G), and the input one. 
Errors refer to the lowest and highest quartile of the distribution.}
 \label{tab:param_retrival}
 \centering
 \begin{tabular}{l c c c c c}
 \hline\hline
  Sample &  m$_{G+G}-$m$_{input}$ & r$_{e,G+G}/$r$_{e,input}$ & n$_{G+G}/$n$_{input}$ & q$_{G+G}/$q$_{input}$ & PA$_{G+G}-$PA$_{input}$ \\
  & (AB mag) &  &  &  & (deg) \\
 \hline
$\mu_{2\arcsec ap} \leq 26.5$ & -0.01$_{-0.02}^{+0.01}$ & 1.00$_{-0.02}^{+0.04}$ & 0.99$_{-0.12}^{+0.07}$ & 1.00$_{-0.01}^{+0.03}$ & 0.08$_{-0.88}^{+0.93}$ \\
$26.5 < \mu_{2\arcsec ap} < 30.5$ & -0.03$_{-0.28}^{+0.07}$ & 1.01$_{-0.23}^{+0.33}$ & 0.95$_{-0.41}^{+0.32}$ & 0.99$_{-0.12}^{+0.11}$ & -0.20$_{-5.94}^{+4.18}$ \\
 \hline\hline
 \end{tabular} 
\end{table*}

\section{Results: MACS1206 the test case}
\label{Sec:results_M1206}

Our test case cluster, MACS1206, is located at RA=12$^h$06$^m$12$^s$.28, 
Dec=-08$^\circ$48\arcmin02\arcsec.4 (J2000), and z=0.44
and it was originally part of the Most Massive Galaxy Clusters survey \citep[MACS][]{Ebeling2001}.
It was codified with morphological class 2, \ie good optical/X-ray alignment and concentric contours \citep{Ebeling2010}
and this relaxed appearance made it a good target for CLASH survey. \citet{Umetsu2012} showed that there is only a small 
offset, \ie 1\arcsec, between the DM peak of mass and the location of the BCG, which coincides also with the X-ray peak 
emission \citep{Ebeling2009}. The excellent agreement between the mass profile of MACS1206 as derived by the 
kinematical analysis \citet{Biviano2013} and the lensing analysis \citet{Umetsu2012} is a further indication that this cluster 
is dynamically relaxed. The global relaxed status of the cluster is also confirmed by the absence of a significant level of 
substructures as found by \citet{Lemze2013}.

We notice that despite this general relaxed condition, MACS1206 displays an elongated 
large-scale structure (LSS) along the NW-SE direction, \citep{Umetsu2012}. This preferential direction is well aligned 
with the position angle (PA) of the BCG and it is traced also by a few infalling groups as revealed by the dynamical analysis 
\citep{Girardi2013}. The cluster has a velocity dispersion $\sigma_{vel}=1087$ km s$^{-1}$ as estimated by the dynamical analysis
of \citet{Biviano2013}, from which we also infer a virial mass M$_{200}$ = 1.41$\times$10$^{15}$ \Msun which is 
in good agreement with the results from weak/strong lensing \citep{Umetsu2012}, and it corresponds to R$_{200}$ = 1.98 
\Mpch.

We run {\it GALtoICL} in the iterated mode on the Rc band image of MACS1206 and use the global best fit model as the benchmark model
to be adapted for the B-band. After obtaining the first temptative global best fit model we allow interactive check and manual 
intervention in case of large residuals. Specifically, for each galaxy showing a high level of residuals we proceed this way:
we checked its z$_{spec}$, if available and consistent with cluster membership, we performed a detailed manual fit and updated the 
global best fit model, while whenever there was not spectroscopic information we masked at different $\sigma_{sky}$ levels. 
When improving the model by manual fitting we generally added a second component to the single S{\'e}rsic model.  
Close-enough initial guesses for each component parameter are important to obtain a reliable fit, thus we took advantage of the
SExtractor+PsfEx softwares combination which allowes spheroid+disk decomposition for each extracted source. The estimated 
{\it MAG\_SPHEROID/DISK, SPHEROID/DISK\_REFF\_IMAGE, SPHEROID/DISK\_ASPECT\_IMAGE, {\rm and} SPHEROID\_SERSICN} values are then used as 
first guess for GALFIT. Tests on simulated galaxies show that manual intervention reduces by 1.5-2.0 times the number of masked pixels 
while providing similar improvement for the residuals in the outermost area of the source segmentation map, \ie where the signal starts to 
blur into sky and small differences in the residuals become important for low SB sources. 

In Fig. \ref{Fig:GALtoICL_results} we show the Rc-band image of the MACS1206 core (left panel), its global best fit model 
(central panel) and the final BCG+ICL map masked down to 1$\sigma_{sky}$ level. The galaxy 
contribution to the light has been removed efficiently and only 4.8\% of the pixels needed to be masked down to 
1$\sigma_{sky}$ level (only 1.4\% when choosing 5$\sigma_{sky}$ level).

In the following we report the results we obtained using the masking down to 3$\sigma_{sky}$ for the SUBARU data which 
corresponds to $\mu_{Rc} \sim 29.3$ mag/arcsec$^{-2}$ at z=0.44.

\begin{figure}
\centering
\includegraphics[width=\hsize]{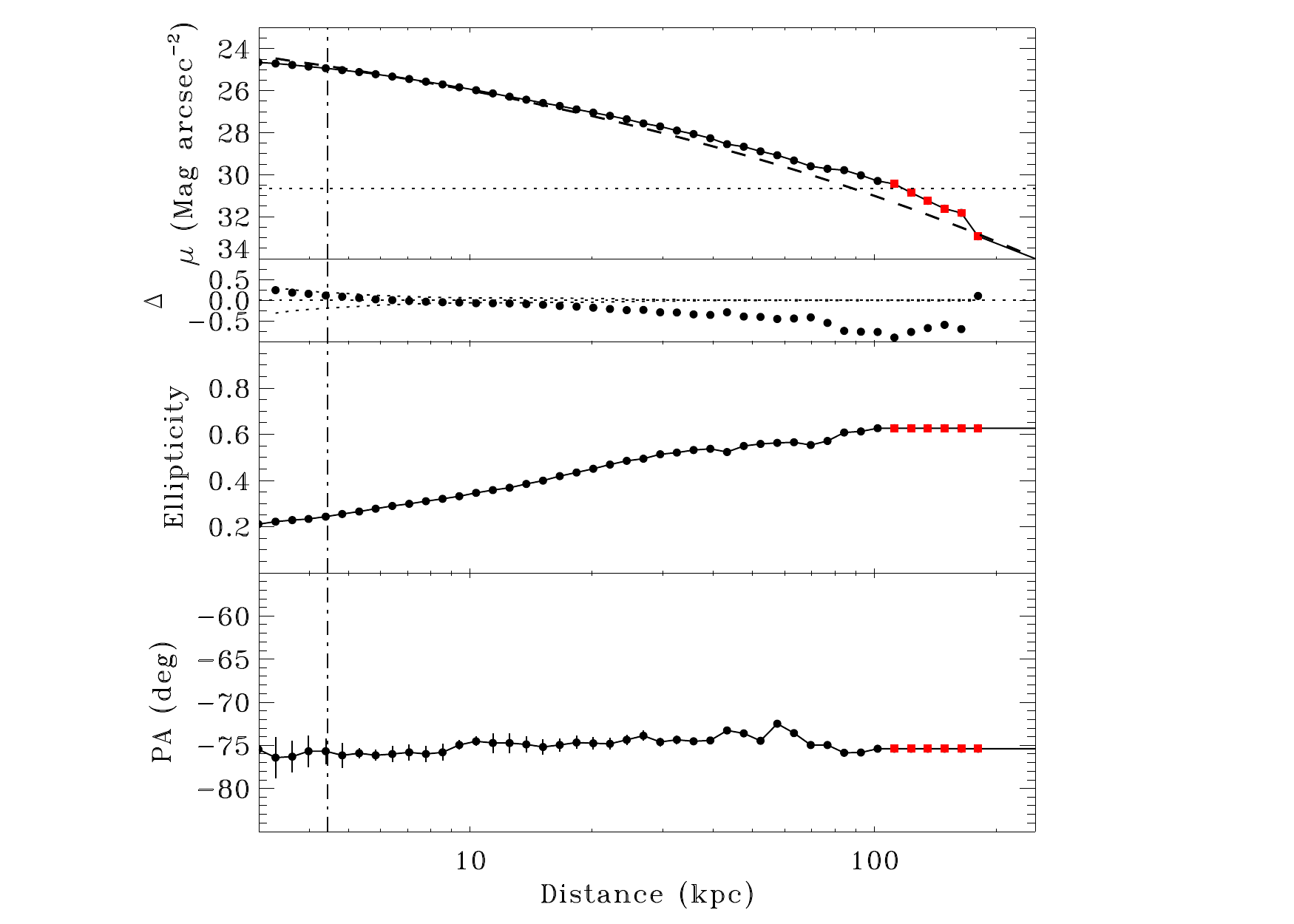}
  \caption{ICL properties: SB profile and residuals to the best fit (top panels), the ellipticity (central panel), and the PA (bottom panel) 
as a function of the distance from the center. The dotted and dashed lines in the top panel refer to the SB at 1$\sigma_{sky}$ level and 
to the best fit model for a de Vaucouleurs profile respectively. Red squares correspond to those points for which the 
isophotal analysis did not converge while the dot-dashed line indicates the psf FWHM limit.}
     \label{Fig:Isophote}
\end{figure}

\subsection{ICL properties}
\label{Subsec:ICLprop}

We performed the classical isophotal analysis of the BCG+ICL using the IRAF \footnote{IRAF is distributed by the 
National Optical Astronomy Observatory, which is operated by the Association of Universities for Research in Astronomy 
(AURA) under cooperative agreement with the National Science Foundation.} task {\it ellipse}. 
We kept the center position fixed and we let the ellipticity and PA vary, Fig. \ref{Fig:Isophote} shows the SB profile and residuals 
to the de Vaucouleurs best fit (top panels), the ellipticity (central panel), and the PA (bottom panel) as a function of the distance 
from the center. We perform a fit of the SB profile with the typical de Vaucouleurs profile, the dashed line in the top panel of 
Fig. \ref{Fig:Isophote} corresponds to the best fit, the dotted line refers to the SB at 1$\sigma_{sky}$ level, and the dot-dashed line 
indicates the psf FWHM limit. Looking at the residuals, it is clear that the r$^{1/4}$ law is a 
poor representation of the data and in the outer region of the BCG, R $\geq$ 40 \kpch, there is an excess of light with respect to the fit. 
This excess of light increases as we move farther away from the center and it is the signature of the ICL. 
At this distance the ellipticity has increased till $\epsilon\ \sim$ 0.55, while the PA has basically a 
constant value of PA $\sim -74 ^{\circ}$ (degrees measured counterclokwise from N direction). In all panels the red squares correspond to those points for which the isophotal 
analysis didn't converge and values are unreliable. We notice that these points are located in the regime where the SB 
reaches the sky level. A close inspection of the BCG+ICL maps reveals an asymmetric elongation of the ICL in the SE 
direction, thus we suppose that in the SE direction we might be able to detect the ICL also at these distances.

\begin{table*}
 \caption{Best fit parameters for different profiles, where 'deVauc' and 'S{\'e}rs' refer to 
 the de Vaucouleurs and S{\'e}rsic profile respectively.}
 \label{tab:BCGproffit}
 \centering
 \begin{tabular}{l c c c c c c}
 \hline\hline
  Profile type &  Mag$_{tot}$ & r$_e$ & n & q & PA & $\tilde{\chi} ^2$ \\
  & (AB mag) & (\kpch) &  &  & (deg) & \\
 \hline
 single deVauc & 18.35$\pm$0.01 & 28.4$\pm$0.3 & 4 & 0.47$\pm$0.01 & -73.42$\pm$0.19 & 19.3 \T\B \\

 single S{\'e}rs & 18.48$\pm$0.00 & 22.4$\pm$0.1 & 3.16$\pm$0.01 & 0.48$\pm$0.01 & -74.24$\pm$0.02 & 34.9 \T\B\\

 single S{\'e}rs ($4<n<8$) & 17.83$\pm$0.01 & 77.1$\pm$1.1 & 6.78$\pm$0.04 & 0.43$\pm$0.01 & -72.74$\pm$0.02 & 2.6 \T\B\\

 \multirow{2}{*}{deVauc+deVauc} & 18.72$\pm$0.07 & 26.3$\pm$1.4 & 4 & 0.51$\pm$0.06 & -79.5$\pm$12.0 &  \multirow{2}{*}{9.6} \T \\
  & 19.41$\pm$0.18 & 37.1$\pm$10.1 & 4 & 0.44$\pm$0.06 & -71.4$\pm$4.6 & \B \\

 \multirow{2}{*}{deVauc+S{\'e}rs} & 19.09$\pm$0.01 & 32.2$\pm$0.07 & 4 & 0.42$\pm$0.01 & -72.33$\pm$0.06 & \multirow{2}{*}{2.5} \T \\
  & 18.13$\pm$0.01 & 138.1$\pm$0.04 & 6.72$\pm$0.04 & 0.43$\pm$0.01 & -74.42$\pm$0.07 & \B \\

 \multirow{2}{*}{deVauc+S{\'e}rs ($n\leq3.99$)} & 19.03$\pm$0.01 & 15.4$\pm$0.07 & 4 & 0.41$\pm$0.01 & -76.35$\pm$0.12 & \multirow{2}{*}{3.0}  \T \\
  & 18.07$\pm$0.01 & 174.7$\pm$1.6 & 3.35$\pm$0.01 & 0.41$\pm$0.02 & -70.11$\pm$0.08 & \B \\

 \hline\hline
 \end{tabular} 
\end{table*} 

\begin{figure*}
\centering
\includegraphics[width=\hsize]{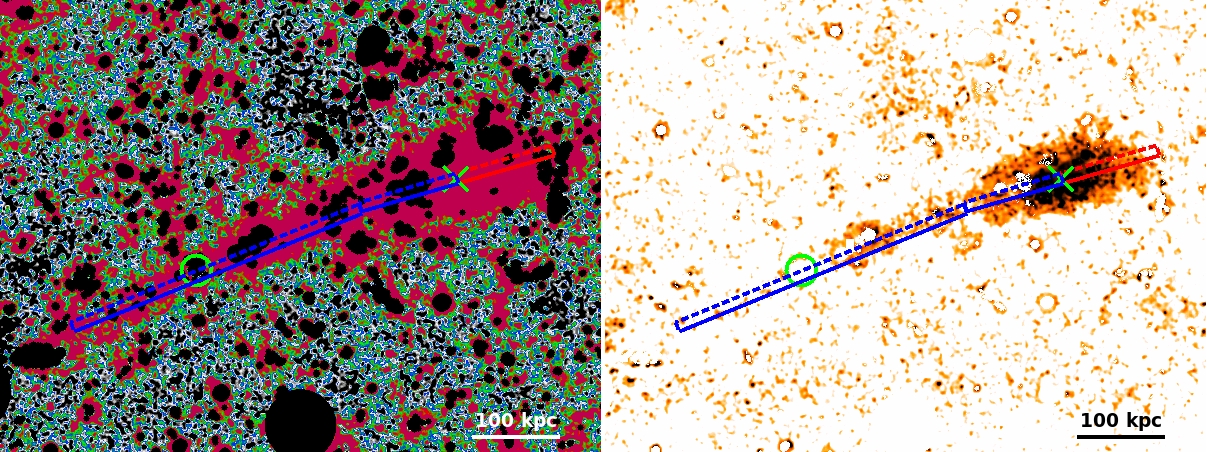}
   \caption{{\it Left panel:} Zoom of the Rc-band BCG+ICL map of MACS1206 smoothed with a Gaussian kernel of 3x3 pixels. We overlaid the
slits along the SE (blue) and NW (red) direction from which we extract the SB profiles. The green cross and circle correspond 
to the location of the BCG and the second brightest galaxy respectively. {\it Right panel:} (B-Rc) color map of BCG+ICL. Slits are overlaid as in the {\it left panel}.}
     \label{Fig:residual_map}
\end{figure*}

\begin{figure*}
 \centering
 \includegraphics[width=\hsize]{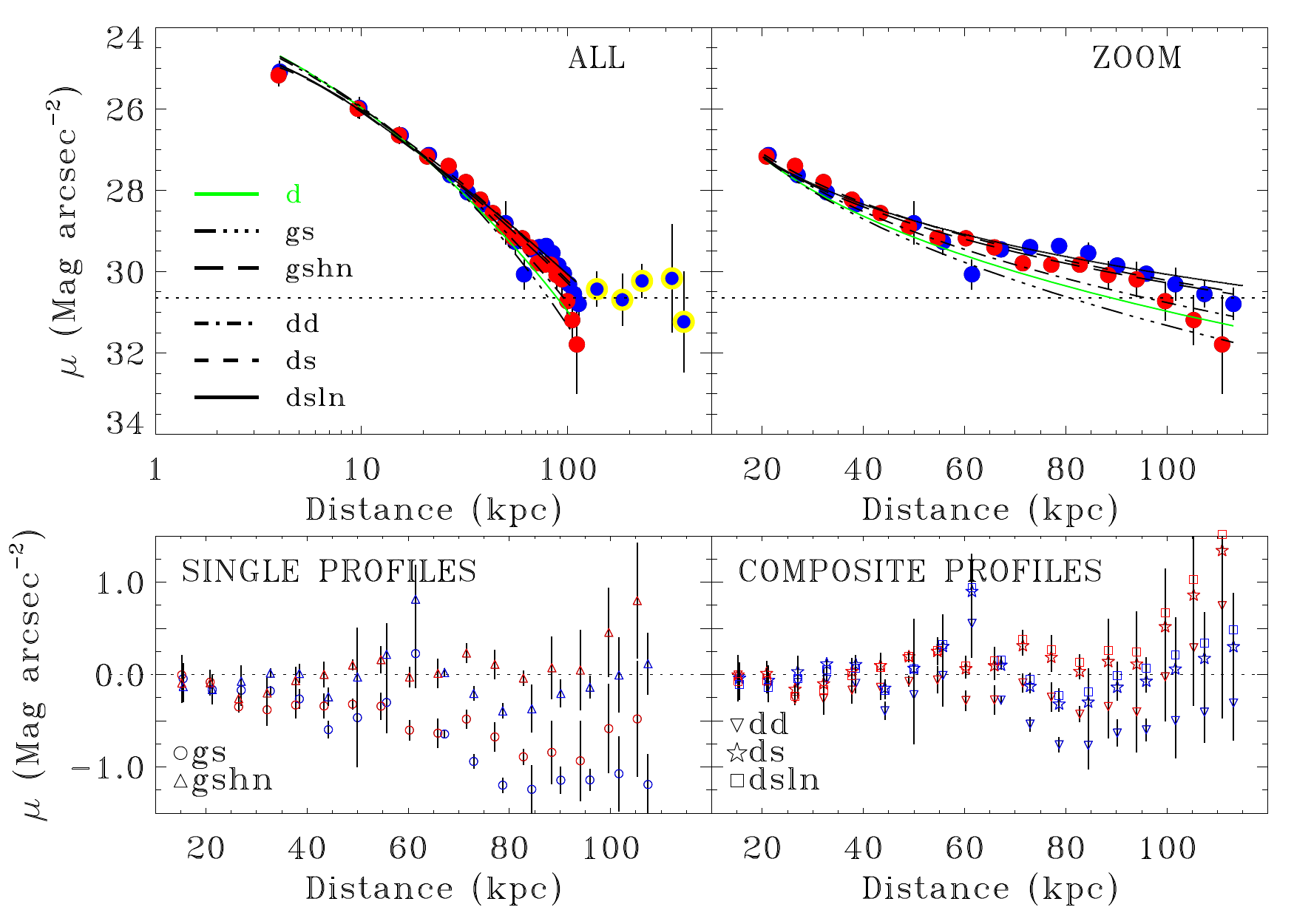}
   \caption{{\it Top Left panel:} SB profile of the Rc-band BCG+ICL map along the SE (blue) and NW (red) directions. 
 Points from the extra slit along the SE direction are highlighted with a yellow circle, while the $\sigma_{sky}$
 level is shown by the dotted line. The generic S{\'e}rsic (gs), generic S{\'e}rsic with high index (gshn), double de Vaucouleurs (dd), 
 de Vaucouleurs plus generic S{\'e}rsic (ds), and de Vaucouleurs plus generic S{\'e}rsic with low index (dsln)
 best fit models are shown by the dot-dot-dot-dashed, long-dashed, dot-dashed, short-dashed, and solid lines respectively.
{\it Top Right panel:} Zoomed version of the SB profile in the radial range $20\leq R \leq100$ \kpch to highlight the asymmetric radial 
 distribution of the SB profile.
{\it Bottom Left panel:} Fit residuals along each direction for the single component profiles, \ie the generic S{\'e}rsic (circles) and 
the generic S{\'e}rsic with high index (triangles). 
{\it Bottom Right panel:} Fit residuals along each direction for the double component profiles, \ie the double de Vaucouleurs 
(upside-down triangles), de Vaucouleurs plus generic S{\'e}rsic (stars), and de Vaucouleurs plus generic S{\'e}rsic with low index 
(squares).}
      \label{Fig:SB_profile_fit}
\end{figure*}

To verify the presence of an asymmetric light distribution, we extract the SB profile from two slits along the PA:
one in the SE direction and the other in the NW direction. In the left panel of Fig. \ref{Fig:residual_map}, we show a smoothed version of 
the BCG+ICL map for the Rc band with the slits overlayed: blue and red colors correspond to the SE and NW direction 
respectively. We located two slits along the SE direction: the main one coinciding with the BCG major axis and an extra 
slit following the ICL elongation towards the second brightest galaxy which is marked with a green circle. We extracted 
the SB profile from each slit and we show it in the top left panel of Fig. \ref{Fig:SB_profile_fit}. 
Points are color coded according to the slit they belong to, the dotted and the green solid lines refer to the sky level and the 
de Vaucouleurs (d) best fit model respectively. To separate between the two slits along the SE direction, we 
highlight with a yellow circle those points obtained from the SE extra slit. The SB profiles along each direction show a similar behaviour within r $\sim$ 60 \kpch, while at larger 
distances the SB profile in the SE direction is systematically above the one in the NW direction. 
Moreover at r $\geq$ 100 \kpch the SB profile in the NW direction blurs into the sky regime, while in the SE 
direction there is still signal. We also detect signal from the extra slit even if it is at sky level. 

Both SB profiles show 
an excess with respect to the single de Vaucouleurs best fit model, so we tried different models to describe the light 
profiles: 1) a generic S{\'e}rsic profile either constraining or not the allowed range for the
S{\'e}rsic index \citep{Oemler1976,Carter1977,Schombert1986,Stott2011}, 2) a double de Vaucouleurs model \citep{Gonzalez2005}, and 3) a composite de Vaucouleurs plus generic S{\'e}rsic profile with either 
free n or within a constrained range of allowed values. In the top left panel of Fig. \ref{Fig:SB_profile_fit} the dot-dot-dot-dashed, 
long-dashed, dot-dashed, short-dashed, and solid lines refer to the generic S{\'e}rsic (gs), generic S{\'e}rsic with high index (gshn), 
double de Vaucouleurs (dd), de Vaucouleurs plus generic S{\'e}rsic (ds), and de Vaucouleurs plus generic S{\'e}rsic with low index 
(dsln) best fit models respectively. The generic S{\'e}rsic best fit profile (n = 3.16) gives even worse results than the 
single de Vaucouleurs one, especially in the outer region where the ICL contribution becomes important. The double 
de Vaucouleurs profile improves the fit even though there is still an excess of light that can not be fit in the outer 
region. This light excess can be better appreciated in the zoomed version of the SB profile in the right panel of Fig. 
\ref{Fig:SB_profile_fit}. Color code and line types are the same as in the left panel, but we show only 
the SB profile at $20\leq R \leq100$ \kpch. On the contrary, both the composite de Vaucouleurs plus generic S{\'e}rsic profiles and 
the single generic S{\'e}rsic profile with $4<n<8$ manage to fit also the light excess at large distances. The de Vaucouleurs 
plus generic S{\'e}rsic with high index profiles provides the best $\tilde{\chi}^2$. The bottom panels show the residulas of single 
component fitted profiles (left) and composite fitted profiles (right). 

In Tab. \ref{tab:BCGproffit} we list the best fit parameters 
for each profile. We notice that both the PA and the ellipticity, $\epsilon=1-q$, show a small range of values among all the adopted
profiles: $-70^\circ\lesssim PA\lesssim-80^\circ$ and $0.59\lesssim\epsilon\lesssim0.49$ respectively. This also suggests that in case 
of a two component profile the BCG and the ICL show a good alignment irrespective of the model choice in agreement with the findings of 
\citet{Gonzalez2005,Zibetti2005}.
In case of a single component fit the effective radius ranges between $\sim20$ \kpch and $\sim80$ \kpch, while when we adopt a composite
profile, the component associated with the BCG has $15 \lesssim r_{e,BCG} \lesssim 32$ \kpch whereas the ICL one is less concentrated and it has 
larger effective radius:  $37 \lesssim r_{e,ICL} \lesssim 175$ \kpch.

As mentioned above we chose to use the Rc band global best fit model as the benchmark model to be adapted for the B-band, 
this enabled us to create a color BCG+ICL map. We degraded the Rc-band image to the same PSF as that of the B 
band, \ie the one with the worst seeing. To transform the PSF of the Rc-band we estimated the kernel function K(r) such
that: PSF$_{Rc band}$(r) $\ast$ K(r) = PSF$_{B band}(r)$, where the symbol $\ast$ denotes a convolution and only
unsaturated stars were used. Sky uncertainties are very challenging in creating color maps, in particular at very low SB
they can significantly affect the final color even if they are very small, \ie at $\mu _{V}$ = 28.5 mag/arcsec$^{-2}$ 
an offset of 1$\sigma_{sky}$ transforms into an uncertainty of $\sim$ 0.2 mags in the (B - Rc) color, while at 2 mag 
brighter the uncertainty is only 0.02. For this reason we rely only on those pixels with $\mu_{V} \leq$ 29.5.

\begin{figure}
\centering
\includegraphics[width=\hsize]{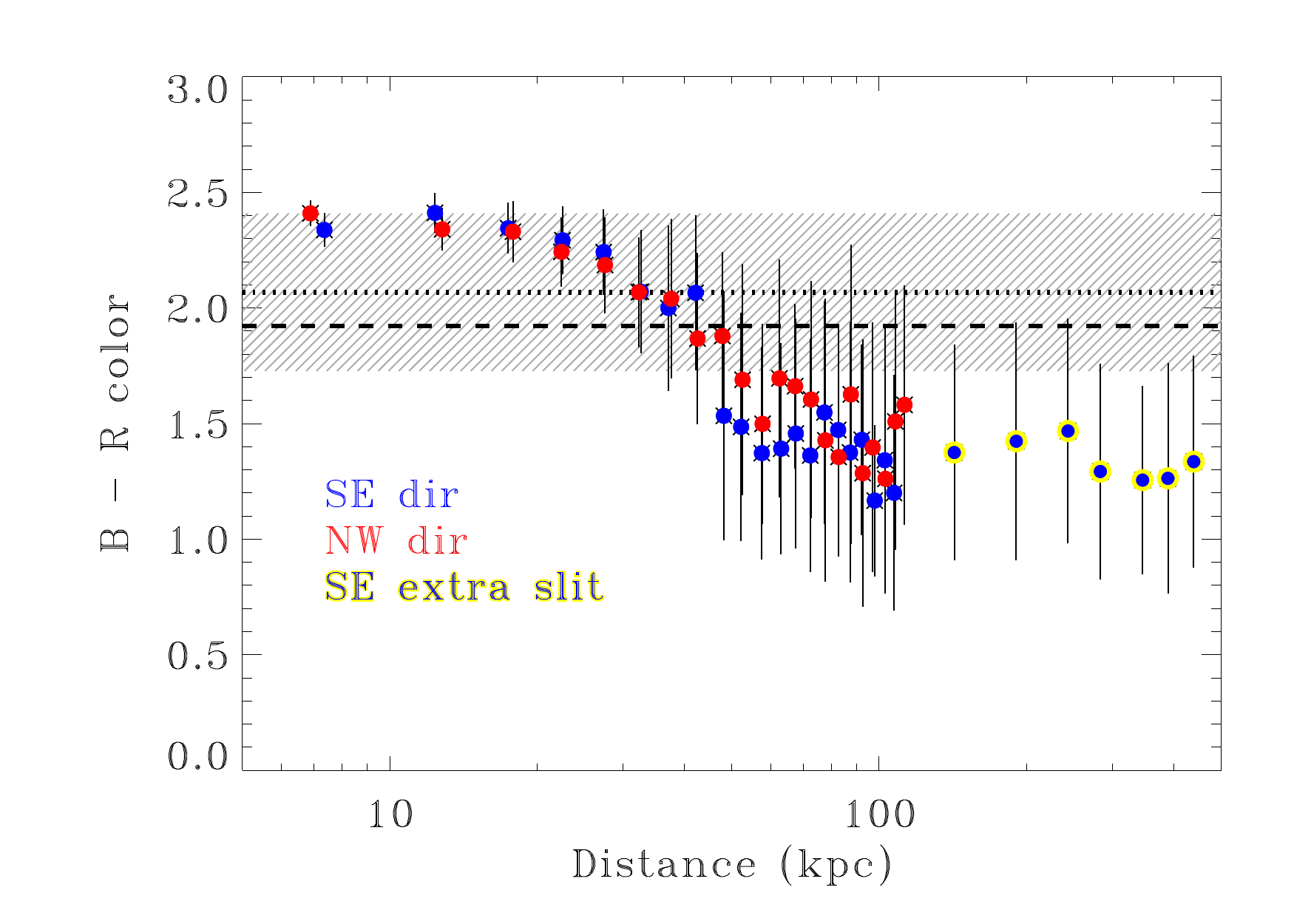}
  \caption{(B-Rc) color profile of the BCG+ICL. Points are color coded as in Fig. \ref{Fig:SB_profile_fit}.
As a reference we overplot the mean (B-Rc) color of cluster member galaxies within R=300 \kpch (dotted line)
and within R$_{500}$ (dashed line). The shaded area correspond to the standard deviation of satellite colors within R=300 \kpch.}
     \label{Fig:ICL_colors}
\end{figure}

In the right panel of Fig. \ref{Fig:residual_map}, we show the (B-Rc) color map 
for the BCG+ICL, the color bar shows exactly the color value that ranges from 2.3 in the very core of the BCG, 
down to 1.5 at distances larger than 50 \kpch. As a reference we overlaid the same slits we used in the SB profile analysis.
At first glance the map shows a color gradient from redder to bluer colors when moving from the 
core of the BCG towards the outer regions which are ICL dominated. We quantified this trend extracting the mean color 
along the slits and in Fig. \ref{Fig:ICL_colors} we show the mean color as a function of the distance from the 
BCG center in bins of 5 \kpch, points are color coded as in the previous plots. The errors correspond to the 
standard deviation of colors in each bin, as expected in the outer regions the large spread in colors shows 
the difficulty to retrieve reliable colors at very shallow SB. There is a bluening trend from the BCG center towards 
outer regions such that the ICL colors tend to be much more similar to those of the outer envelope of the BCG rather 
than its central region. This is consistent with previous results, \eg \citep{Zibetti2005,Rudick2010}. However the 
BCG+ICL is reliably detected only out to r=50 \kpch in the B band, \ie 2$\sigma$ detection, thus the bluening trend 
is milder if we consider only the safe detection region. A linear fit to the color profile out to r=50 \kpch returns a 
slope of -0.16 $\pm$ 0.12 in $\Delta$(B - Rc )/$\Delta \log(r)$ which is compatible with zero gradient or very weak 
negative gradient. As a reference we overplot the mean (B-Rc) color of cluster member galaxies within R=300 \kpch (dotted line)
and within R$_{500}$ (dashed line). The shaded area correspond to the standard deviation of satellite colors within R=300 \kpch
which is approximately the same for satellites within R$_{500}$. We note that BCG+ICL colors within the safe detection region, \ie 
r$\sim$50 \kpch, are in good agreement with those of the satellite galaxies residing in the core of the cluster.

\begin{figure}
\centering
\includegraphics[width=\hsize]{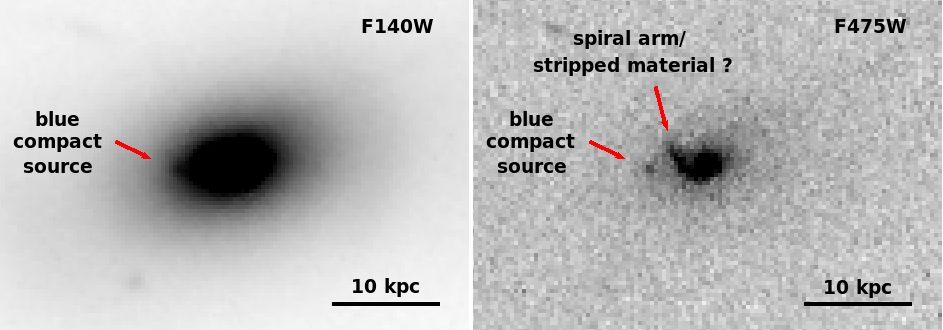}
  \caption{{\it Left panel}: F140W image of the BCG center, the green cross point is located at the BCG 
  center and the presence of a small source in the SE direction is highlighted by a red arrow. 
  {\it Right panel}: same as above but for the F475W filter. In this bluer filter the source is well separated from the 
BCG center. This filter highlights also the presence of a sort of spiral arm in the very center of the BCG in the same 
direction of the compact blue source, see text for details.}
     \label{Fig:blue_source_center_BCG}
\end{figure}

The color profile along the two directions is in good agreement within the error bars but we note that 
the innermost point, \ie r $\leq$ 10 \kpch, in the SE direction tend to be bluer than the corresponding one along the NW direction, 
though within 1$\sigma$. This bluening is confirmed by the presence of [OII] emission in the BCG spectrum obtained by our team with 
FORS2 as part of the program 090.A-0152(A) \citep[see][]{Grillo2013}. This [OII] emission line was already noted by 
\citet{Ebeling2009} and it was interpreted as an evidence in favour of MACS1206 being a CC cluster. 
However a careful inspection of HST data reveals the presence of both a compact source and an inner core spiral arm at $\sim$ 1\arcsec, 
\ie $\sim$ 6 \kpch, which are completely blended to the BCG center in the SUBARU data due to their pixel scale. 
Both these features are embraced in the spectrum aperture and may be responsible for the [OII] emission. 
Left panel of Fig. \ref{Fig:blue_source_center_BCG} shows the HST F140W image of the BCG 
center \citep[see][for the description of HST image observation and data reduction]{Postman2012,Koekemoer2011}, 
the green cross point is located at the BCG center and the presence of a small source in the SE direction is 
highlighted by a red arrow. In the right panel we show the same region but for the F475W filter, whose transmission curve
brackets the [OII] emission redshifted at the cluster redshift. In this bluer filter 
the blue compact source is well visible and separated from the BCG center. This filter highlights also the presence of 
a sort of spiral arm in the very center of the BCG extending only in the SE direction. Given that this 
structure is present only in one direction it is more probable to be a residual of stripped material.

Our data may suggest that this [OII] emission can be associated to the blue compact source and/or peculiar features 
blended with the BCG core emission, but we can not exclude the presence of a moderate/weak CC. Whether MACS1206 is a CC or not 
is far beyond the purpose of this paper, thus we refer the reader to Appendix \ref{appen:CC} for a brief discussion of this point.
For the sake of completeness, we should mention the possibility of the blue compact source being a fore- background source, 
while the spiral arm seems connected to the BCG center.

\subsection{ICL contribution to the total mass budget}
\label{Subsec:ICLfrac}

We determined the BCG+ICL fraction as a function of the cluster-centric radius. We extracted 
the total flux within a set of circular apertures from both the BCG+ICL map 
and the total members map. To create the total members map
we need to assign membership to each source in the field of view and we
rely on the cluster membership as described in Sect. \ref{Subsec:cl_membership}.
We mask all the light contribution from fore- and back-ground galaxies down to 1$\sigma_{sky}$, 
while bright stars were identified using the CLASS\_STAR parameter of SExtractor, \ie
CLASS\_STAR$>0.98$, and we create an ad-hoc mask to ensure spikes coverage.

\begin{figure*}
   \centering
    \includegraphics[width=\hsize]{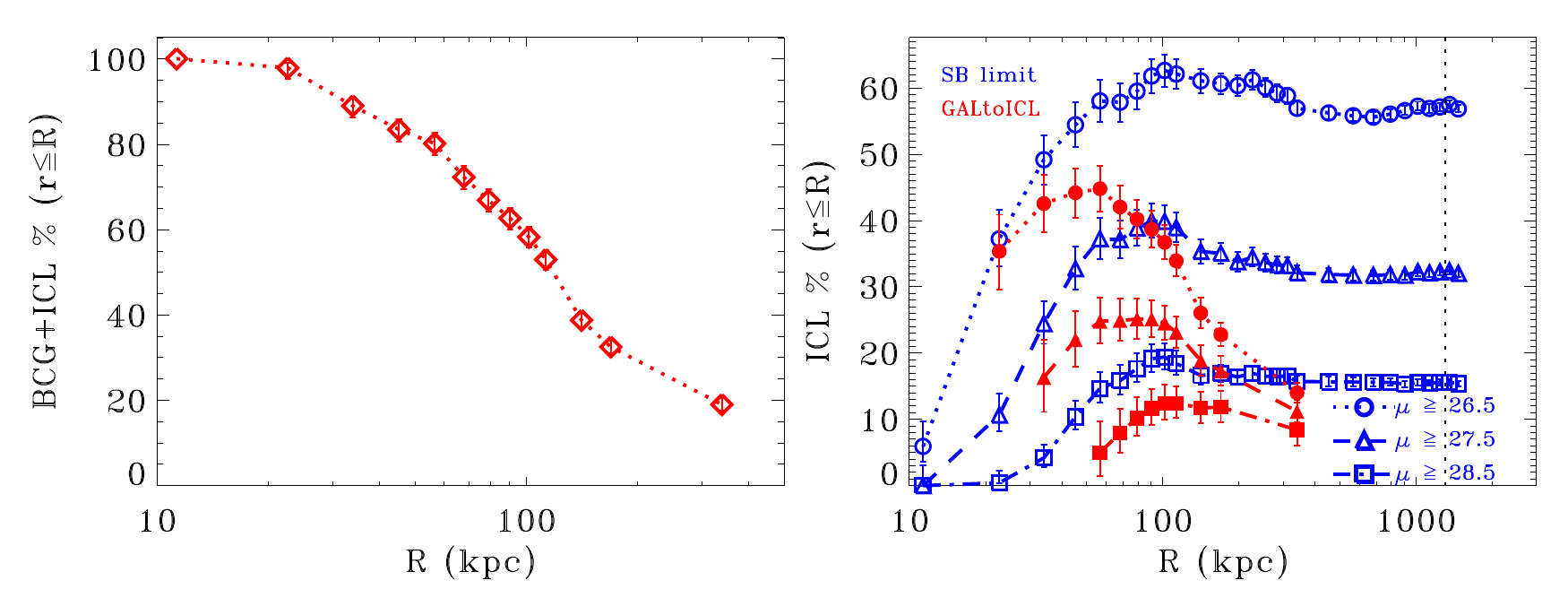}
      \caption{{\it Left panel:} BCG+ICL contribution to the total cluster light within each circular 
apertures of radius R as derived from the residual map obtained using the {\it GALtoICL} code. {\it Right panel:} ICL fraction
as a function of the cluster-centric distance for different surface brightness levels and different ICL measurement methods. 
Empty symbols refer to the SB limit method while filled ones refer to the {\it GALtoICL} code. Circles, triangles and squares correspond 
to $\mu_{Rc}=26.5,\, 27.5,\, \mathrm{and}\, 28.5$ mag/arcsec$^2$ surface brightness levels respectively. The dotted line at R$\sim$1300 kpc indicates R$_{500}$.}
         \label{Fig:gala_vs_SBlim}
   \end{figure*}

In the left panel of Fig. \ref{Fig:gala_vs_SBlim} we show the BCG+ICL 
contribution to the total 
cluster light within each circular aperture of radius R. 
Error bars are estimated in a similar 
way as in \citet{Djorgovski1984}: we divide each aperture into eight 
sections and estimate the total flux in each sector. The error bars represent 
the rms of total flux in each sector thus taking into account the possible lumpiness 
of light distribution in each aperture.

We note that at 100 \kpch the BCG+ICL contributes more than 50\% 
while at R$\sim$350\kpch it drops down to $\sim20$\% of the light within 
that circular aperture. This BCG+ICL percentage is also confirmed by the analysis of
the dark matter profile decomposition performed by \citet{Grillo2013} at a similar radial distance.

In our approach we extract BCG+ICL maps because it is not trivial to 
distinguish between the two components and we decide to avoid any a priori 
separation. However we can quantify the ICL contribution by combining the 
de Vaucouleurs + S{\'e}rsic profile parameters that best fit the SB profile of the BCG+ICL, 
see Sect. \ref{Subsec:ICLprop}, and a proper M/L conversion.
Our (B-Rc) color analysis shows that the ICL color tend to be similar to that
of the BCG outer envelope, \ie it can be treated as a red/passive source.
To derive the M/L conversion for the ICL we then determine the best fit of
the relation between the stellar masses of red cluster member galaxies, 
\ie 2.0$\leq$(B-Rc)$\leq$2.5 and the total Rc magnitude of their best 
fit model we obtained with GALAPAGOS:
\begin{equation}
 \log(M/M_\odot) = (19.43\pm0.94) - (0.41 \pm 0.04) \times Rc_{tot\,mag}
\end{equation}
where we use stellar masses by \citet{Annunziatella2013}, \ie obtained by SED fitting
using the MAGPHYS software \citep{daCunha2008}, based on the 2007 version of the BC03 models 
\citep{Bruzual2003,Bruzual2007} with Chabrier IMF \citep{Chabrier2003} and assuming a 
set of exponentially declining star formation histories and random bursts 
superimposed to them. 
Applying this relation to the total Rc magnitude of the de Vaucouleurs plus generic S{\'e}rsic best fit model,
we obtain M$_{ICL}=(9.9\pm3.8)\times10^{11}$M$_{\odot}$ and $M_{BCG}=(4.0\pm2.1)\times10^{11}$M$_{\odot}$.

By summing all the galaxy stellar masses of cluster members down to $\log(M/M_\odot)=9.5$, \ie the stellar mass
completeness limit corresponding to 23 mag in Rc band \citep[][see text for details]{Annunziatella2013}, out to R$_{500}$
and that of the BCG as obtained using the above calibration we obtain the total 
stellar mass of the cluster, M$_{*,\,500}=(1.7\pm0.7)\times10^{13}\,M_\odot$. 
Error bars on M$_{*,\,500}$ are obtained by summing in 
quadrature the typical galaxy stellar mass error and errors from standard bootstrap technique.
The critical radius R$_{500}$ is determined using the NFW profile for M$_{200}=(1.4\pm0.2)\times10^{15}$M$\odot$
and c$_{200}=5.8\pm1.1$ as obtained by the lensing analysis of \citet{Umetsu2012} and
we get R$_{500}$=1.3 Mpc which means M$_{500}=1.0\times10^{15}$M$_{\odot}$.
The ICL contains 5.9$\pm$1.8\%,of the stars within R$_{500}$, while the BCG+ICL contribution to M$_{*,\,500}$ is 8.2$\pm$2.5\%. 
As a further check we estimated the light contained in the de Vaucouleurs + Sersic best fit model, \ie in the BCG+ICL components,
out to R$_{500}$, and we summed the light of each member galaxy out to R$_{500}$, rather modelling them, to obtain the total cluster 
light out to R$_{500}$. The corresponding BCG+ICL and ICL fractions are 6.3$\pm$0.6\% and 4.3$\pm$0.2\% respectively. These values are 
in good agreement with those obtained converting the BCG+ICL total magnitudes into stellar masses within the errorbars.

The corresponding contribution of stars, f$_{*}$ to the total mass of the cluster,
taking into account also the ICL contribution, 
is then (M$_{*,\,500}$+M$_{ICL}$)/M$_{500}$=0.0177$\pm$0.006.
We should also remind that the total galaxy stellar mass within R$_{500}$ is affected by projection effects that tend 
to increase its value. If we consider a spherical cluster having MACS1206 values for M$_{200}$ and c$_{200}$ and extending out to 
3$\times$R$_{200}$, then the 2D projected mass within R$_{500}$ is 1.56$\times$M$_{500}$.Taking into account this 
projection effect, 
than M$_{*,\,500,\,{\it deproj}}=1.18\times10^{13}$M$_{\odot}$ corresponding to 
f$_{*,\,{\it deproj}}=0.0116\pm0.006$, 
where we have excluded the BCG from the correction as it lies in the center of the cluster. 

\subsection{Comparison with the surface brightness method}
\label{Subsec:ICL_cfr_SBlim}

We now compare these results with those obtained using a different definition of the ICL.
We determine the ICL fraction by applying the same approach 
of many works in the literature \citep[][and references therein]{Krick2007,Burke2012}:
choosing an arbitrary SB cut-off level below which pixels are masked and counting 
all the light above this level as the ICL. This ICL definition is a very 
naive way to separate galaxy light and ICL, but it is the most suitable 
definition from the operational point of view and for comparison purpose.
Moreover we will be able to explore advantages/disadvantages of each method and 
to reveal possible systematics.

We produced ICL maps using SExtractor segmentation maps: we set-up the 
THRESH\_TYPE parameter to {\it absolute} mode and we choose three different 
SB cut-off thresholds: 26.5, 27.5 and 28.5 mag/arcsec$^2$.
This way the sources are extracted only down to each SB level and the segmentation 
maps correspond to the galaxy light to be masked.
In the ICL maps those pixels associated to either a source counterpart in the 
segmentation maps, or stars, fore- and back-ground galaxies, or sky areas
were masked. All the remaining pixels are considered as ICL.

\begin{figure*}
   \centering
      \includegraphics[width=\hsize]{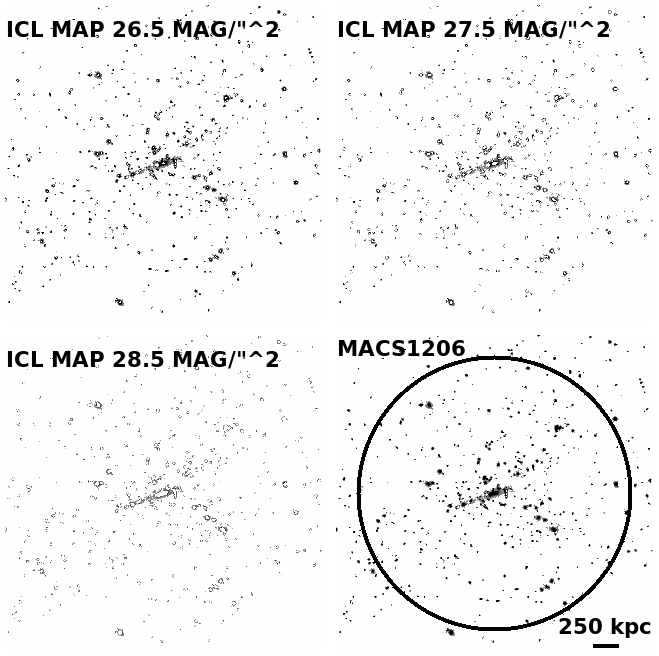}
      \caption{Images show only cluster members light below given 
surface brightness levels which is considered as ICL. The surface brightness
limits correspond to $\mu_{Rc}$ = 26.5, 27.5, 28.5 mag/arcsec$^2$ and total 
cluster light from top left to bottom right. The black circle in the bottom right panel corresponds to R$_{500}$}
         \label{Fig:cfr_diff_SBlim}
   \end{figure*}

In Fig. \ref{Fig:cfr_diff_SBlim} we show the Rc-band ICL map down to 26.5, 27.5 and 
28.5 mag/arcsec$^2$ and the total cluster light map from top left to bottom 
right. These images show the same asymmetric light distribution along the 
SE-NW direction in the proximity of the BCG as we found with the {\it GALtoICL} code.

These images have only a display purpose, to quantify 
the ICL fraction we sum-up all the flux contained in circular apertures out 
to R$_{500}$ for each image in Fig. \ref{Fig:cfr_diff_SBlim}.

In the right panel of Fig. \ref{Fig:gala_vs_SBlim} we show the ICL 
contribution to the total light for each SB level. Blue empty circles,
triangles, and squares refer to to 26.5, 27.5, and 28.5 mag/arcsec$^2$ 
surface brightness levels respectively while the dotted line indicates R$_{500}$.

The fraction of ICL shows a common trend among all SB levels: it has a steep increase
from the core out to R$\sim$100 kpc where it reaches its maximum, then it shows a plateau. 
Given that the BCG+ICL fraction as obtained with the {\it GALtoICL} 
code accounts for more than 50\% of the light at R$\sim$100 kpc and then it drops quite rapidly, then 
the plateau trend at larger radii can only be justified as light contribution from the 
other member galaxies. As a further confirmation, we masked the BCG+ICL map with a circle
centered on the BCG and a radius corresponding to the typical distance at which the BCG
SB profile reaches 26.5, 27.5, and 28.5 mag/arcsec$^2$, \ie R$\sim15,\, 30,\, \mathrm{and}\, 50$ kpc.
We then extracted the light in the same aperture as before and we determine its contribution 
to the total light. This is shown by the filled symbols in the right panel of 
Fig. \ref{Fig:gala_vs_SBlim}, different symbols correspond to different SB masking levels as 
before. We notice that at large radii, \ie $\sim300$\kpch, the ICL contribution drops to 10-15\% depending on the adopted SB limit.
This suggests that most of the ICL is concentrated in the close surroundings of the BCG, while at larger distances the ICL 
constribution is not significant.

By comparing the ICL fraction as obtained from the {\it GALtoICL} code and the SB limit method we 
note that even at small distances, \ie at R$\sim$50 kpc, there is a significant difference 
between them. Moreover the general trend of increasing ICL fraction out to R$\sim$80-100 kpc 
is still present, but then at larger radii the ICL fraction drops down to a small percentage 
instead of showing an almost constant value. This reinforces the idea that the SB limit 
method can be contaminated by the light coming from the outer regions of cluster member 
galaxies. Despite this, the SB limit method is still the easiest way one can use to compare the observational 
results to the expected values from the simulations or to other observational studies.
Thus we applied the SB limit method for $\mu_{Rc} = 28.87, 29.87$, \ie the $\mu_{V}(z=0) = 26.5, 27.5$
SB levels transformed into Rc-band at z=0.44, see Sect. \ref{Subsec:efficency}. The corresponding
ICL fraction at R$_{500}$ are 12.5 $\pm$ 0.6\% and 4.7 $\pm$ 0.4\% respectively. 

These ICL fraction are based on our deepest and best ICL detection filter, the Rc band, but we have 
multi band imaging of this cluster thus we decided to determine the first ICL SED to measure ICL stellar mass fraction. 
We use the Rc-band masks for each SB levels as reference masks on the others bands, 
\ie B, V, and Ic (having adapted masks to differences in seeing conditions among different bands), we then mask stars, 
fore-, and back-ground in each band according to their detections 
down to 1$\sigma_{sky}$ level. Finally we extracted the light which survived to the masking and that
is associated to the member galaxies according to SExtractor segmentation maps within R$_{500}$
in each band.

\begin{figure}
   \centering
   \includegraphics[width=\hsize]{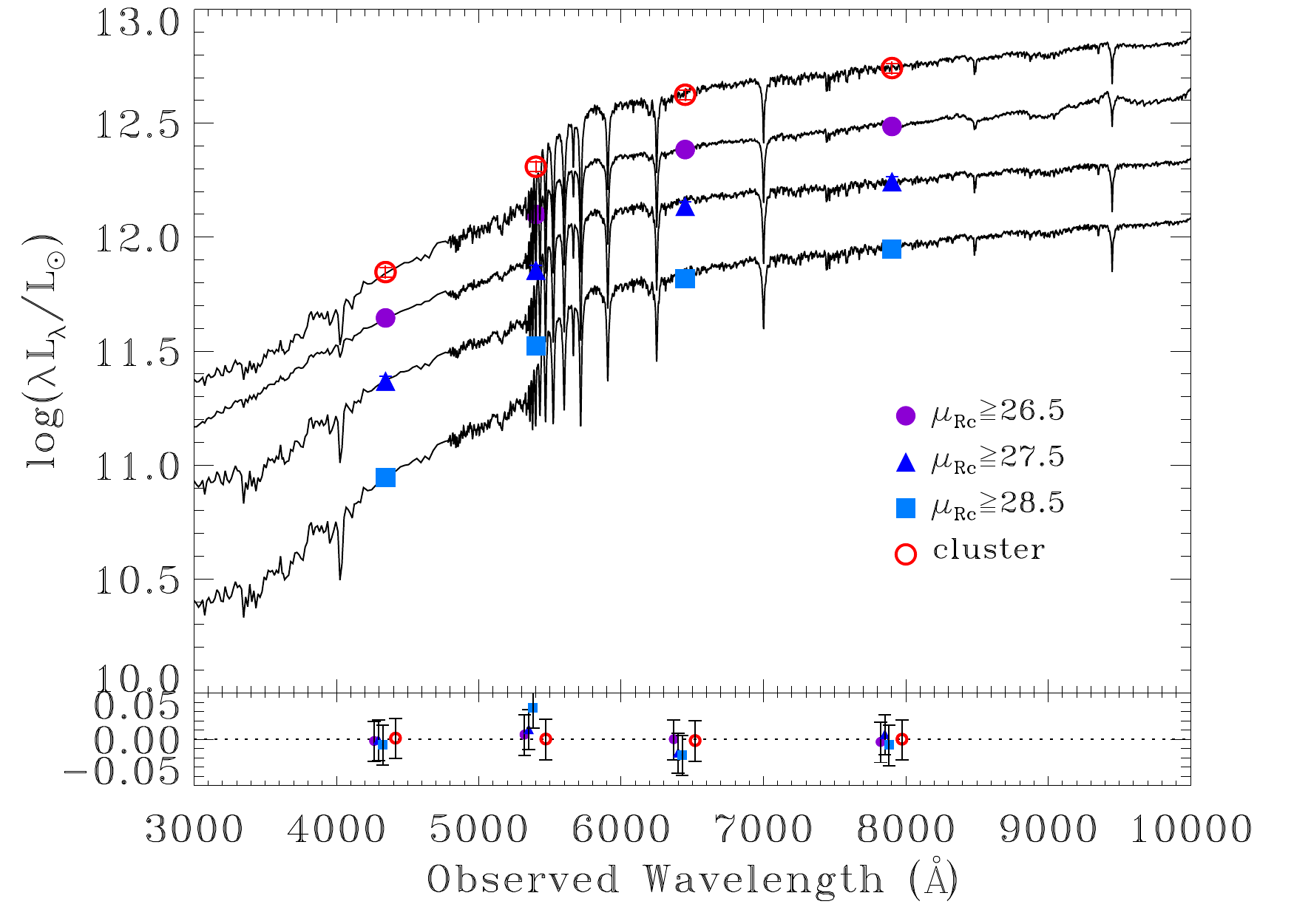}
      \caption{{\it Top panel:} SED of the total cluster light within R$_{500}$ (red empty circles) and that 
of the ICL within R$_{500}$ for different SB limits: $\mu_{Rc}(z=0.44)$ = 26.5, 27.5, 28.5 mag/arcsec$^2$ 
(violet filled circles, blue filled triangles, and cyan filled squares respectively).
{\it Bottom panel:} residuals between the observed fluxes in each band and those obtained using the SEDs best fitting
models for each SB level.}
         \label{Fig:ICL_SED_fit}
   \end{figure}

In the top panel of Fig. \ref{Fig:ICL_SED_fit} we show the SED of the total cluster (red empty circles) and that 
of the ICL for different SB limits: $\mu_{Rc}$ = 26.5, 27.5, 28.5 mag/arcsec$^2$ (violet filled circles, blue 
filled triangles, and cyan filled squares respectively). We performed a fit to these SEDs using the software
MAGPHYS and the black solid lines in the top panel of Fig. \ref{Fig:ICL_SED_fit} represent the SED best 
fitting models for the cluster and ICL.
In the bottom panel of Fig. \ref{Fig:ICL_SED_fit} we plot the residuals between the observed fluxes 
in each band and those obtained using the SEDs best fitting models for each SB level.
The ICL mass fraction obtained from the SED fits range between 20\% and 55\% depending on the choosen SB
level and qualitatively in agreement with the {\it SBlimit} values.
We did not repeat the same exercise for the $\mu_{Rc}(z=0.44) = 28.87,29.87$ because the corresponding B and V band
masks cover already the whole galaxies, \ie at these SB levels we reach the sky regime.

\section{Discussion}
\label{Sec:discussion}

We developed an automated method to create BCG+ICL maps and we measured a diffuse intracluster component in MACS1206.
We confirm previous findings on general ICL properties: 1) a composite profile best fits the data \citep{Gonzalez2005,Zibetti2005}, 
though we find that a de Vaucouleurs plus S{\'e}rsic profile provides a better fit than a double de Vaucouleurs one,
2) BCG and ICL position angles agreee within few degrees 
\citep{Gonzalez2005,Zibetti2005} and both are in agreement with the global cluster elongation and its filament 
\citep{Umetsu2012,Girardi2013}, and 3) ICL colors agree with those of the outer envelope of the BCG
\citep{Zibetti2005,Krick2006,Pierini2008,Rudick2010}. 

Disentangling the BCG component from the ICL is one of hardest task when studying the diffuse light and for this reason we 
preferred to create BCG+ICL maps. However in order to quantify the ICL properties and its contribution to the total cluster light 
we shall separate it from the BCG. We tried different profiles, either single or composite ones by combining the de Vaucouleurs and 
the S\'ersic profiles. Ellipticies and PA show a small range of values both in case of a single and composite profiles, while the 
effective radius show a wider range depending on the adopted profile. In case of a single component fit the effective radius ranges 
between $\sim20$ \kpch and $\sim80$ \kpch, while when we adopt a composite profile, the component associated with the BCG has 
$15 \lesssim r_{e,BCG} \lesssim 32$ whereas the ICL one is less concentrated and it has larger effective radius: 
$37 \lesssim r_{e,ICL} \lesssim 175$.

 \noindent \citet{Ascaso2011} analyzed a sample of BCGs at a similar redshift, 
they fitted them with both a single de Vaucouleurs and a generic S\'ersic profile and they find $<r_{e,de Vauc}> = 19 \pm 10$ \kpch 
and  $<r_{e,Sers}> = 23 \pm 15$ \kpch. Their mean effective radii are in good agreement with our results if we consider that MACS1206 has 
a higher X-ray luminosity than that of \citet{Ascaso2011} sample, \ie L$_{X,0.1-2.4 keV}=24.3\cdot10^{44}$ erg s$^{-1}$, and that 
larger BCGs are located in more massive clusters. Similarly, \citet{Stott2011} find $<r_{e,de Vauc}> = 27 \pm 2$ \kpch 
and  $<r_{e,Sers}> = 57 \pm 16$ \kpch at higher redshift, \ie z$\sim1$. 
Concerning the effective radius of the outer component for the double de Vaucouleurs fit, we find a small radius when compared
to \citet{Gonzalez2005}. Their mean effective radii of the ICL component is $\sim$160 kpc though 20\% of their sample have 
r$_{e,ICL}<50$ kpc, thus small ICL effective radius are not ruled out. We should also consider that 
our double de Vaucouleurs profile is not able to properly fit the outer component, see residuals in bottom panel of 
Fig. \ref{Fig:SB_profile_fit}, thus it might be that we are also underestimating R$_e$. 
On the contrary, the effective radius of the outer component for the de Vaucouleurs + Sersic profiles has a larger value, 
$\sim 140$ kpc.

The most peculiar feature of the ICL in MACS1206 is its asymmetric 
radial distribution: there is an excess of ICL in the SE direction. Peculiar streams of ICL are  
supposed to last only $\sim$1.5 times their dynamical timescale in the cluster according to simulations  
\citep{Rudick2009} because of disruption by cluster tidal field. More generally the streams found in the cluster core
live only $\tau_{ICL\, survival} \leq 1$ Gyr due to the strong tidal fields they are subject to. Thus the galaxy/ies from which this material 
has been stripped away should have interacted with the BCG no later than a Gyr ago. Moreover the ICL enhancement along the SE  
direction extends out to the second brightest galaxy which is classified as an {\it H$\delta$ red} galaxy, 
\ie poststarburst galaxies (PSBs). The spectral properties of PSB galaxies can only be reproduced by either models of galaxies in a 
quiescent phase soon after a starburst ( $\tau_{PSB}  \leq$ 1.5 Gyr) or by models where a regular star formation has been halted in 
an abrupt way \citep{Poggianti1999}. Recently \citet{Pracy2013} showed that H$\delta$ equivalent width radial profiles
in local PSBs can be reproduced by merger simulation at even shorter ages after the peak of the starburst: 0.2-0.75 Gyr. 
The ICL survival timescale and that of PSBs are in good agreement, thus the ICL stream along the SE direction can be interpreted as the 
stars stripped from the second brigthest galaxy which has crossed the cluster, sunk to the center, and interacted with the BCG. 
We note that the second brightest galaxies is aligned with the ICL extra slit PA along the SE direction, see Fig. \ref{Fig:residual_map}.
The dynamical analysis of MACS1206 has hilighted the presence of a preferential direction which is traced by both the {\it passive}
and {\it H$\delta$ red} galaxies with $PA_{{\it H\delta/Passive}} \sim 110^{\circ}$ 
(measured counter–clock–wise from north) \citep{Girardi2013}. 
Matching our BCG/ICL PA estimates, we find $101^{\circ} \leq PA_{BCG/ICL} \leq 109^{\circ}$ which is similar to this preferential 
direction, thus suggesting a further connection between the ICL and the infalling direction of the PSBs population.
This scenario is also supported by the presence of an elongated large scale structure
(LSS) around the cluster whose major axis runs along the NW-SE direction, $15^{\circ} \leq PA_{LSS} \leq 30^{\circ}$ measured north of west
\citep{Umetsu2012}. Matching our PA estimates to the same reference system as \citet{Umetsu2012} we find $11^{\circ} \leq PA_{BCG/ICL} \leq 19^{\circ}$, 
depending on the assumed BCG+ICL best fit profile. Thus both the BCG and the ICL are oriented along the same axis as that of the LSS,
this holds also when comparing the ellipticity of the LSS and of the BCG+ICL.
As a consequence the BCG of MACS1206 should have experienced a strong interaction that dates back to at least 
$\tau_{past\,merger} \leq 1.5$ Gyr ago, this interaction might involve also the second brightest galaxy and it may 
has occured along the preferential NW-SE direction.

Both observation and simulations suggest that short-lived major mergers can produce a significant fraction of the ICL 
\citep{Burke2012,Burke2013,Murante2007,Laporte2013,Contini2013}. If we consider the extreme case of the second brightest galaxy 
merging into the BCG of MACS1206, we can determine the dynamical friction timescale and compare it with the light travel time to z=0. 
If the former is shorter than the latter, then we can roughly estimate the 2$^{nd}$ brightest galaxy contribution to the ICL at the 
end of the merging process. The dynamical friction timescale for a galaxy of mass M$_{gal}$ at a given initial radius R$_{in}$ that 
spirals into the center of the cluster potential well on a circular orbit with velocity V$_c$ is given by Eq. \ref{eq:tau_df}
\citep{Binney1987}:

\begin{equation}
 \tau_{df}=1.17 \cdot \frac{R_{in}^2 V_c}{\ln(\varLambda) G M_{gal}}
\end{equation}
\label{eq:tau_df}

\noindent
where $\ln(\varLambda)$ is the Coulomb logarithm, $\ln(\varLambda)\sim \frac{b_{max}V_c^2}{GM_{gal}}$. In the cluster core the impact
parameter, $b_{max}$, is roughly 100 kpc, the typical circular velocity is 
V$_c \sim \sqrt{2}\cdot\sigma \sim \sqrt{2}\cdot1100 \sim 1500$ km s$^{-1}$, where we used the velocity dispersion obtained by 
\citet{Biviano2013}, and the 2$^{nd}$ brightest galaxy has 
M$_{gal} \sim M_{gal,*}/f_{baryon,gal} \sim 10^{11.5}/0.05 \sim 6.3 \cdot 10^{12}$, where we used the galaxy stellar mass obtained by
\citet{Annunziatella2013} and the typical baryon fraction of early-type galaxies \citep{Hoekstra2005,Jiang2007}. Thus, 
$\ln(\varLambda)\sim 2.2$ and $\tau_{df} \sim 2.7$ Gyr, given the projected radial distance between the 2$^{nd}$ brightest galaxy and
the BCG, R$_{in} \sim 300$\kpch. \citet{Nath2008} find similar dynamical timescales values for a massive galaxy (M$_{gal}=3 \times 10^{12}$M$_\odot$) 
embedded in a rich cluster (M$_{cl}=10^{15}$M$_\odot$) at a similar initial radius. Equation \ref{eq:tau_df}
is based on strong approximation, \ie circular orbit and point-like object.  \citet{B2008} take into account the effect of an 
extended object with different orbital parameters on the $ \tau_{df}$ estimate and find that standard approximation tend to shorten the
dynamical friction timescale. They also provide a fitting formula to determine the merging timescale due to dynamical friction as a 
function of both the satellite to host halo mass ratio and the satellite orbital properties, see their Eq (5). If we consider the host 
halo as mainly composed by the BCG+ICL, M$_{host}=10^{12.1}$, and we assume the same baryon fraction as for the 2$^{nd}$ brightest galaxy, 
then our mass ratio is $M_{sat}/M_{host} = 10^{11.5}/10^{12.1} \sim 0.25$. Allowing the initial circularity and the initial orbital 
energy parameter to vary in the same validity range as \citet{B2008}, \ie 0.33-1.0 and 0.65-1.0 respectively, we obtain
$1.0 \lesssim \tau_{merge,df} \lesssim 6.0$ Gyr with a $\langle\tau_{merge,df}\rangle \sim 2.6$.
The light travel time to z=0 is $\sim 4.6$ Gyr, thus there is enough time for the 2$^{nd}$ brightest galaxy to merge into the BCG, if this 
is the case.

The fraction of ICL coming from galaxies that merged with the BCG ranges between 5\% to 30\% for the most massive clusters depending on
the simulation set-up \citep{Murante2007,Puchwein2010,Laporte2013,Contini2013}. If the 2$^{nd}$ brightest galaxy is going to merge with
the BCG, then it will release $1.6-9.5 \times 10^{10}$M$_\odot$ to the ICL by z=0. This corresponds to $\sim1-10$\% of the ICL at z=0.44
and this increase is well within the errorbars, similar consideration can be made in terms of f$_{ICL}$ which would become 
$\sim5.9-6.4$\%.

\begin{figure*}
   \centering
   \includegraphics[width=\hsize]{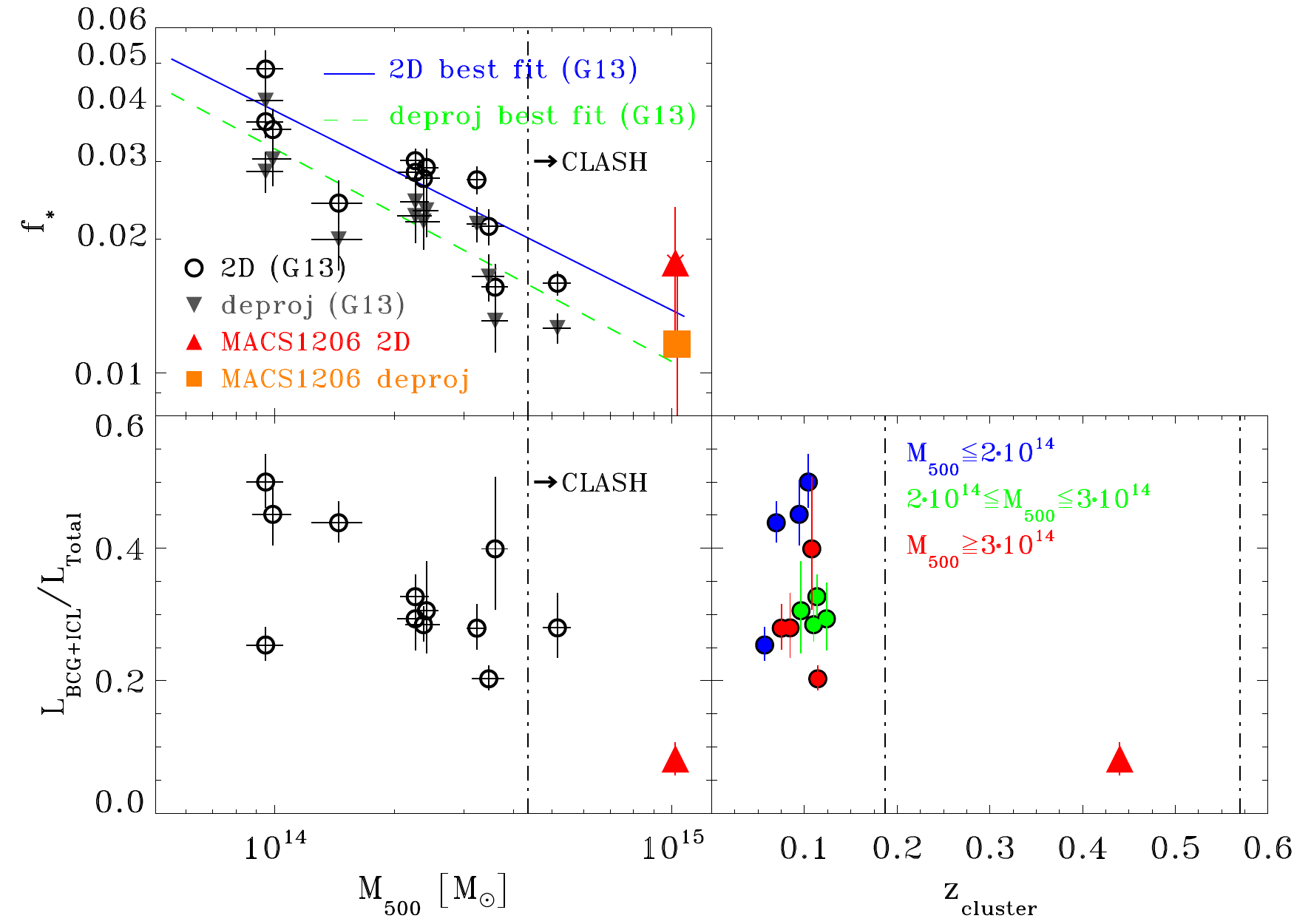}
      \caption{{\it Top panel:} Stellar baryon fraction as a function of M$_{500}$ for both MACS1206 and the cluster sample of 
\citet[][G13 hereafter]{Gonzalez2013}. 
(Orange square) Red triangle refers to the (de-)projected f$_*$ for MACS1206, while (upside-down grey triangles)
open circles refer to the (de-)projected G13 sample. The (green dashed) blue solid line correspond to
the (de-)projected best fit relation from G13 while the dot-dashed line indicates the predicted cluster mass M$_{500}$
lower limit for the CLASH sample (see text for details).
{\it Bottom left panel:} BCG+ICL fraction of light/mass within R$_{500}$ as function of cluster mass for both MACS1206 and the G13 cluster sample, 
symbols/lines as above .
{\it Bottom right panel:} BCG+ICL fraction of light/mass within R$_{500}$ as function of cluster redshift. G13 sample is color coded according 
to their M$_{500}$. Blue, green, red circles correspond to M$_{500} \leq 2\times10^{14}$M$_\odot$ ,
$2\times10^{14} \leq $M$_{500} \leq 3\times10^{14}$M$_\odot$, and M$_{500} \geq 4\times10^{14}$M$_\odot$ respectively.}
         \label{Fig:cfr_G13}
   \end{figure*}

We quantified the mass contribution of the BCG+ICL to the stellar cluster mass within the critical radius R$_{500}$ as $\sim8$\%,
this value is in good agreement with the general trend of decreasing BCG+ICL mass (light) fraction with increasing cluster mass
\citep[][G13 hereafter]{Lin2004,Gonzalez2007,Gonzalez2013}. For comparison purpose in the bottom left panel of Fig. \ref{Fig:cfr_G13} we show
BCG+ICL fraction of light (mass) within R$_{500}$ as a function of cluster mass for both MACS1206 and the \citet{Gonzalez2013} 
cluster sample, red triangle and open circles respectively. \citet{Gonzalez2013} provides BCG+ICL luminosity fractions while
we estimate the mass BCG+ICL fraction. According to \citet{Cui2013} luminosity-weighted and mass-weighted ICL fractions are in good 
agreement especially at the high cluster mass end of their sample, \ie the ratio of luminosity to mass fractions at 
M$_{500} \sim 10^{15}$ M$_\odot$ is consistent with 1 when AGN feedback is taken into account. The dot-dashed line indicates the predicted cluster mass M$_{500}$
lower limit for the CLASH sample according to the M-T$_X$ best fit relation of \citet{Mahdavi2013} and to the CLASH cluster selection 
T$_X \geq 5 keV$. We note that the expected cluster mass range covered by the CLASH sample will fill the lack of observational data at 
the high mass end, thus allowinging this kind of study on a wider cluster mass range and with a well constrained total cluster mass estimate.
On top of this, the CLASH/VLT sample will also span a wider range in cosmic time and we will be able to study the BCG+ICL contribution to the 
cluster stellar mass disentangling between halo mass and redshift dependences, if any. 
In the bottom right panel of Fig. \ref{Fig:cfr_G13} we show the BCG+ICL fraction as a function of redshift, the G13 sample is color coded 
according to their M$_{500}$: Blue, green, red circles correspond to M$_{500} \leq 2\times10^{14}$M$_\odot$ ,
$2\times10^{14} \leq $M$_{500} \leq 3\times10^{14}$M$_\odot$, and M$_{500} \geq 4\times10^{14}$M$_\odot$ respectively.

We notice that the ICL stellar mass (light) of MACS1206 represents $\sim72$ (70)\% of that of the BCG+ICL assuming our best fit model 
parameters and the adopted mass to light conversion. 
 Though using a different composite profile, we obtain similar results to \citet{Gonzalez2005} with a large percentage of the
light residing in the outer component, the one associated to the ICL.
 As a consequence, the ICL contribution on small scales is very important, though on larger scales it becomes less significative. 
This is clearly shown in the right panel of Fig. \ref{Fig:gala_vs_SBlim} once we adopt a SB threshold on our BCG+ICL maps, 
\ie red points, on the contrary applying the same SB limit to the original image shows a plateau of the ICL fraction 
at large radii. This highlights the systematic error in the ICL contribution estimate depending on the adopted 
method: light from the outer envelopes of member galaxies can significantly affect the ICL fraction when using the SB 
limit method. This effect is larger at lower SB limits, but even at the higher SB limit the estimated ICL fraction is twice 
that obtained with the {\it GALtoICL} method. Once again we stress the importance of removing all the light from galaxy members 
that can affect the real ICL contribution. Unfortunately the {\it SBlimit} method is the best way to compare results among observational works 
and simulations. We find good agreement between our ICL fractions at Rc-band SB levels corresponding to 
$\mu_V (z=0) \ge 26.5$ mag/arcsec$^2$ and those expected from simulations. For a cluster with the same M$_{500}$ as MACS1206, 
\citet{Cui2013} estimates ICL fraction at R$_{500}$ of 10-20\% and 5-10\% for $\mu_{V}(z=0) = 26.5 \, \mathrm{and} \, 27.5$ respectively depending on the 
adopted simulation, \ie with either gas cooling, star forming, and supernova feedback or including
AGN feedback, thus showing good agreement with our results.
\citet{Rudick2011} simulated clusters with a smaller mass range, still if we consider their 
most massive cluster B65, M$_{200}$=$6.5 \times 10^{14}$ M$_{\odot}$, the ICL fraction for $\mu_{V}(z=0)=26.5$
is nearly 12\% within $1.5 \times R_{200}$, see left panel of their Fig. 3. Given that they claim 
only a smaller increase in the ICL fraction within R$_{500}$, these values are in good agreement with our results.
A direct comparison with observational works is less trivial due to different ICL enclosing radius or lack of cluster total mass 
information. For instance \citet{Feldmeier2004} finds ICL fraction of $\sim$10 (2)\% above $\mu_{V}(z=0)=26.5 (27.5)$  mag/arcsec$^2$
for a set of clusters located at z$\sim$0.17. These values are in good agreement with our ICL fraction of $\sim$12 (4)\% at Rc-band 
SB levels corresponding to $\mu_V (z=0) \ge 26.5 (27.5)$ mag/arcsec$^2$, thus suggesting a lack of evolution in the ICL fraction with 
cosmic time. This result agrees with the absence of strong variation in the amount of ICL between z=0 and z=0.8 reported by
\citet{Guennou2012} and other authors \citep{Krick2007}. However we should remind that this comparison is regardless of the cluster 
total mass and/or ICL enclosing radius. On the contrary, we should mention that most of the simulation studies report a 
significant increase of the ICL with time. Irrespective of the formation redshift of the ICL, simulations show that roughly 60-80\%
of the ICL present at z=0 is built up at z$<$1 \citep{Murante2007,Rudick2011,Contini2013}.
Both simulation and observation suggest that part of the ICL origins from tidal disruption 
 of intermediate-mass galaxies as they interact with the BCG or the other most massive galaxies in the cluster 
\citep{Willman2004,Murante2007,Coccato2011,Martel2012,Giallongo2013}. 
This scenario is supported by the analysis of environmental dependence of the galaxy mass function of MACS1206
\citep[see][]{Annunziatella2013}.

We estimate the total star contribution to the baryon fraction and both our projected and de-projected f$_{*}$ are 
in good agreeement with the results of the recent analysis of \citet{Gonzalez2013} where they also considered the 
effects of projection. More generally our values agree with previous studies and the general trend of low f$_{*}$
for the most massive clusters \citep{Andreon2010,Zhang2011,Lagana2011,Lin2012,Gonzalez2013}. In the top panel of
Fig. \ref{Fig:cfr_G13} we show f$_*$ as a function of M$_{500}$ for both MACS1206 and the cluster sample of \citet{Gonzalez2013}. 
(Orange square) Red triangle refers to the (de-)projected f$_*$ for MACS1206, while (upside-down grey triangles)
open circles refer to the (de-)projected G13 sample. The (green dashed) blue solid line correspond to
the (de-)projected best fit relation from G13 while the dot-dashed line indicates the predicted cluster mass M$_{500}$
lower limit for the CLASH sample as in the bottom left panel. We note that our estimate of f$_*$ is in excellent agreement with the 
expectation from the best fit relation of G13. Once again we stress that at completion CLASH/VLT will enlarge the baseline of
the f$_*$-M$_{500}$ relation with the advantage of a well constrained cluster total mass.

Adding the gas fraction f$_g$=0.144$\pm$0.025 as estimated by \citet{Ettori2009} to the stellar component, we obtain
 the total baryon fraction f$_b$=0.156$\pm$0.026,
to be compared with f$_b$=0.167 (0.154) as expected from WMAP7 
(PLANCK) results \citep{Planck2013,Komatsu2011}. The comparison with PLANCK results is less straightforward due to
different cosmological parameters which have a strong impact as shown by \citet{Gonzalez2013}. Our total baryon 
fraction is 7\% below the expected value but well within 1$\sigma$. Generally speaking this result is in agreement with the trend of increasing (decreasing) 
f$_{gas}$ (f$_*$) with cluster total mass, thus supporting the idea of a less efficient star formation at the high end of the 
cluster mass function \citep[][and references therein]{Andreon2010,Zhang2011,Lagana2011,Lin2012,Gonzalez2013}.

\section{Summary and Conclusions}
\label{Sec:conclusions}

In conclusions we have developed an authomated method to extract BCG+ICL light
maps in a refined way: {\it GALtoICL}. Applying this technique to MACS1206:

\begin{enumerate}
 \item We have highlighted the presence of an extra component, \ie the ICL, when studying the SB profile of
the BCG. This component appears to be asymmetric in radial distribution and we interpret it as an
evidence of a past merger. We have linked the ICL properties to those of the cluster substructures
and this way we have reconstructed the most recent cluster assembly history.
  
  \item We have estimated the BCG+ICL mass fraction and the (de-) projected f$_*$ of MACS1206 to be 
in good agreement with recent literature results suggesting a lowering in star formation efficency
at higher cluster masses.

  \item We have estimated the sole ICL contribution with two different methods, {\it GALtoICL} and the {\it SBlimit} methods, 
and compared their results. The {\it SBlimit} method provide ICL fractions systematically larger than those obtained with the
{\it GALtoICL} method due to member galaxies, other than the BCG, light contamination. The {\it GALtoICL} method removes this 
contamination by fitting simultaneously galaxies, thus providing safe ICL detection and it also highlights the 
presence of features/plumes in the ICL. As a con, the {\it GALtoICL} method is much more time consuming compared to 
simpler methods such as the SB limit definition and it can only be applied to small field of view.

  \item Based on the {\it SBlimit} method, we have obtained the first temptative ICL global SED. The ICL mass fraction
we obtained by the SED fitting are in qualitative good agreement with those simply obtained by fluxes in the single reference 
broadband filter Rc. 

\end{enumerate}

The high-quality dataset, the new refined ICL detection method, and the comparison of different ICL detection methods are the most 
striking novelties of this work. 
Deep multiband photometry allowed us to securely detect the ICL at a relatively high redshift, z=0.44, while 
the spectroscopic information allowed us to select cluster members, determine their masses down to $\log$(M/M$_\odot$)=9.5
and thus obtain an accurate estimate of the cluster stellar mass, BCG+ICL stellar mass, and f$_*$. The wide spectroscopic 
dataset also permit to associate the ICL properties to the dynamical analysis of MACS1206 and thus reconstruct its assembly 
history. While a single data point can not give statistical relevance to our results and/or allow to draw strong conclusions, 
at completion the CLASH/VLT survey will provide a high quality dataset over a wide redshift range, thus enabling us to constrain
both the role of the ICL in the baryon budget and the f$_*$-M$_{500}$ relation.

This work has also highlighted the importance of a common definition of ICL to allow comparison among both observational and 
numerical works. Simple ICL definition such as the {\it SBlimit} method might be easier to compare but they do not retrieve the real ICL
properties because of contamination effects. 

\begin{acknowledgements}
We thank the anonymous referee for constructive comments that help us to improve 
the manuscript.
VP is grateful to Monaco, P., Murante, G., and De Grandi, S. for useful discussion and comments.
VP acknowledges the grant ''Cofinanziamento di Ateneo 2010'' and financial support from
PRIN-INAF2010 and MIUR PRIN2010-2011 (J91J12000450001). 
WC acknowledges a fellowship from the European Commission's
Framework Programme 7, through the Marie Curie Initial Training
Network CosmoComp (PITN-GA-2009-238356), supports from ARC DP130100117 
and from the Survey Simulation Pipeline (SSimPL; {\texttt{http://ssimpl-universe.tk/}}).
AF acknowledges the support by INAF through VIPERS grants PRIN 2008 and PRIN 2010. 
Support for AZ is provided by NASA through Hubble Fellowship grant \#HST-HF-51334.01-A awarded by STScI.
\end{acknowledgements}

\bibliographystyle{aa}
\bibliography{ICL_Presotto14_accepted}

\appendix
\section{BCG [OII] emission line}
\label{appen:CC}

Our team obtained a medium resolution spectrum of the BCG with FORS2 as part of the program 090.A-0152(A) \citep[see][]{Grillo2013}. 
We measure the [OII] Equivalent Width (EW) from an aperture of $\sim1.5$\arcsec, 
\ie 9 \kpch diameter, around the peak emission of the BCG flux calibrated spectrum: EW$_{OII}=-4.9\pm3.2$\AA.
This corresponds to L$_{[OII]}=7.4 \pm 4.8 \times10^{40}$ \ergsec, having multiply the EW by the flux density of the 
best-fitting SED at 3727\AA. The level of our [OII] 
emission line detection is very low, in contrast to what is expected for strong/moderate cool core (CC) \citep{Crawford1995} and 
in agreement with normal BCG showing no/low [OII] emission \citep{Samuele2011}. This [OII] emission line was already noted by 
\citet{Ebeling2009} and it was interpreted as an evidence in favour of MACS1206 being a CC cluster. \citet{Ebeling2009} also note 
that the [OII] emission was at a much lower level than typically observed in large CC clusters, thus flagging MACS1206 as a moderate 
CC cluster. Using a different parameter, also \citet{Baldi2012} classify MACS1206 as a CC cluster even if the temperature profile is 
approximately constant around kT $\sim$ 10 keV. This kind of temperature is very high as compared to typical CC central temperatures, 
\ie 3-4 keV \citep{Finoguenov2001} and it also has a too low central metallicity, \ie 0.25 
\citep[][see also the ACCEPT web site \footnote{http://www.pa.msu.edu/astro/MC2/accept}]{Cavagnolo2009}, with respect to typical CC.
\citet{Cavagnolo2009} also estimated the central cooling time, $\tau_0\sim1$Gyr, and the central entropy, K$_0\sim70$keV cm$^2$, of MACS1206. 
These values are borderline between the absence of CC and the presence of a weak CC according to the multi-parameter analysis of 
\citet{Hudson2010}.

\end{document}